\documentclass[reprint,superscriptaddress,amsmath,amssymb,aps,twocolumn]{revtex4-1}
\usepackage{graphicx}
\usepackage{bm}
\usepackage{amsmath,amssymb,amsfonts}
\usepackage{epsfig}
\usepackage{epstopdf}
\usepackage{dcolumn}
\usepackage{grffile}
\usepackage{verbatim}
\usepackage{mathrsfs}
\usepackage{xr}
\usepackage{extarrows}
\usepackage[normalem]{ulem}
\usepackage[colorlinks=true,linkcolor=blue,citecolor=blue, urlcolor=blue]{hyperref}

\def\bk{{\bm{k}}}
\begin{document}
\title{Roton Superconductivity from Loop-Current Chern Metal on the Kagome Lattice}
\author{Zhan Wang}
\affiliation{Department of Physics, Boston College, Chestnut Hill Massachusetts 02467, USA}
\author{Keyu Zeng}
\affiliation{Department of Physics, Boston College, Chestnut Hill Massachusetts 02467, USA}
\author{Ziqiang Wang}
\email{wangzi@bc.edu}
\affiliation{Department of Physics, Boston College, Chestnut Hill Massachusetts 02467, USA}
\date{\today}
\begin{abstract}
Motivated by the evidence for time-reversal symmetry (TRS) breaking in {\em nonmagnetic} kagome metals $A$V$_3$Sb$_5$ ($A=$ K, Rb, Cs), a novel persistent electric loop-current (LC) order has been proposed for the observed charge density wave (CDW) state. The LC order and its impact on the succeeding superconducting (SC) state are central to the new physics of the kagome materials. Here, we show that the LC order fundamentally changes the nature of the pairing instability and the resulting SC state, giving rise to a novel topological superconductor, dubbed as a roton superconductor. In the single-orbital model on the kagome lattice, the CDW state with LC order is a Chern metal near van Hove filling with a partially filled Chern band hosting three Chern Fermi pockets (CFPs). Cooper pairing of quasiparticles on the CFPs generates three SC components coupled by {\em complex} Josephson couplings induced by the TRS breaking LC. Due to the discrete quantum geometry associated with the 3-fold rotation and sublattice permutation, a small LC produces a significantly large Josephson phase that drives the leading SC instability to the roton superconductor where the relative phases of the three SC components are locked at 120$^\circ$, forming an emergent vortex-antivortex lattice with simultaneous pair density modulations. The roton superconductor supports topological chiral edge states carrying nonzero electric currents along domain walls and the sample boundary. A unique property of the roton superconductor is the low-energy fractional $1/3$ vortex and antivortex pairs excitations associated with supercurrent phase slips. Remarkably, the bound states of 3 Cooper pairs, each residing on a different CFPs, are immune to such internal chiral phase fluctuations and form coherent charge-6$e$ Cooper pair triplets. We discuss these extraordinary SC properties in connection to recent experimental evidence for TRS breaking chiral SC state in kagome superconductors, exhibiting pair density modulations, zero-field SC diode effect, and charge-6e flux quantization. These findings are also relevant for the interplay between the orbital-driven quantum anomalous Hall and SC states in other systems such as the graphene and transition metal dichalcogenide based quantum materials.

\end{abstract}

\maketitle

\section{Introduction}
The prototypical correlated and topological quantum state is a Chern insulator exhibiting the quantum anomalous Hall effect (QAHE). This state requires the spontaneous breaking of time-reversal symmetry (TRS) and was introduced as the Haldane phase on the honeycomb lattice in the presence of a microscopic charge current on the atomic scale~\cite{1988Haldanea}. The QAHE has been observed in magnetic topological materials~\cite{2013Xue}. When carriers are introduced into a Chern insulator, a partially filled Chern band with a nontrivial Chern number can be created, giving rise to a gapless Chern metal. For a narrow Chern band, the Fermi energy can be much smaller than the correlation energy, and the interaction-driven gapped stable phases exhibiting the fractional QAHE have been proposed~\cite{2011Bernevig,2011DasSarma,2011Mudry,2011Sheng,2011Wen} and recently observed~\cite{2023Xud,2023Caia,2023Shan,2023Xuc,2024Jua}. The Chern metal can also be unstable by becoming a superconductor. However, superconductivity from a partially filled Chern band has remained largely unexplored.

The recent discovery of {\em nonmagnetic} kagome superconductors $A$V$_3$Sb$_5$ ($A=$ K, Cs, Rb)~\cite{2019Ortiz, 2020Ortiza, 2024Ortiz} offers a surprising route to the new physics on the kagome lattice. The normal state of the kagome superconductors is a triple-$\bm{Q}$ charge density wave (CDW) metal, which is highly unconventional and intertwined with additional symmetry breakings beyond the broken lattice translation symmetry. Central to the revelation is the debate surrounding evidence for TRS breaking revealed by scanning tunneling microscopy (STM)~\cite{2021Jiang, 2021Hasan}, laser STM and piezo-magnetic responses~\cite{2023Xingb}, muon spin rotation ($\mu$SR)~\cite{2022Guguchia, 2024Guguchia, 2021Zhao}, optical Kerr rotation~\cite{2022Wangnl, 2022Wuliang, 2023Farhang, 2023Saykin, 2024Wang}, circular dichroism~\cite{2022Wuliang}, magneto-chiral transport~\cite{2022Moll, 2024Moll}, and magnetotropic susceptibility measured by tuning fork resonators ~\cite{2025Jiao}. In the absence of spin-related magnetism, a density wave of persistent electric loop-current (LC) has been conjectured for the metallic CDW state~\cite{2021Jiang} and received considerable theoretical attention~\cite{2021Hu, 2021Nandkishore,2021Neupert,2022Zhoub,2022Fernandes,2023Zhou,2023Leea,2023Kontani,2023Sushkov,2024Kee,2024Wu,2025Scammell,2025Vanderbilt}. 

In the simplest one-orbital, 3-band model on the kagome lattice at van Hove (vH) filling, the $2\times2$ CDW with LC order gaps out the nested Fermi surface and produces a topological orbital Chern insulator, a kagome lattice analog of the Haldane phase for the QAHE on the honeycomb lattice~\cite{1988Haldanea}. It is realizable in concrete model studies including the electronic interactions~\cite{2023Zhou,2023Kontani,2024Kee,2024Wu}. The proximity to vH filling introduces extra carriers to the LC CDW Chern insulator and leads to a partially filled Chern band with Chern Fermi pockets (CFPs) carrying concentrated Berry curvature~\cite{2022Zhoub}, which has been shown to be a stable LC Chern metal driven by intersite Coulomb interactions~\cite{2023Zhou}. The staggered loop-currents are persistent charge currents in the quantum ground state and generate TRS breaking nonrelativistic kinetic magnetism.

Here we study the nature of the superconducting (SC) states emerging from the LC Chern metal with a partially filled Chern band on the kagome lattice.
We show that the LC CDW order fundamentally changes the nature of the pairing instability and the resulting SC state,
producing an unprecedented chiral topological vector superconductor.
Beyond revealing the new physics of electronic correlation and topology, we develop concrete insights into the theoretical studies of the kagome superconductors near vH filling~\cite{2012Li, 2021Thomale, 2022Zhoub, 2022Kontani, 2023Leea, 2024Wangqh}. Our findings also provide a plausible explanation for the recent experimental evidence for TRS breaking chiral SC state in kagome superconductors, exhibiting pair density modulations~\cite{2024Yin,2024Yinb} and primary pair density waves~\cite{2021Chen, chenhui-surface}, zero-field SC diode effect~\cite{2024Le,2025Wang}, as well as charge-6e flux quantization supporting higher-charge SC correlations in CsV$_3$Sb$_5$ thin film ring structures~\cite{2024Ge} that has attracted substantial theoretical interests~\cite{2022Zhoub,2022Patrick,2024Hud,2023varma-wang,2024Wub,2025ZhangGM}.
\begin{figure}
	\includegraphics[width=\columnwidth]{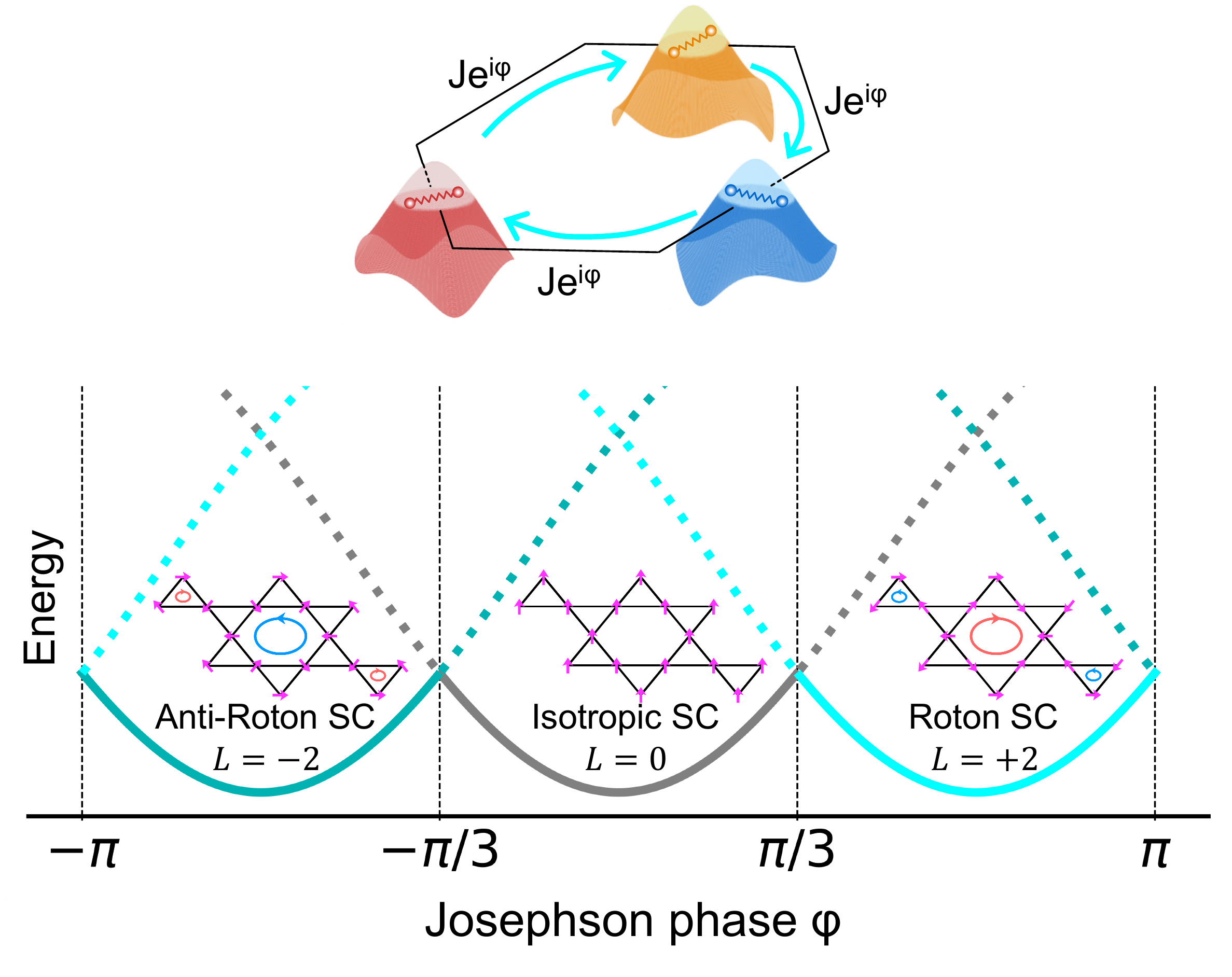}
	\caption{{\bf Schematic SC phase structure from LC Chern metal}. Upper panel: a three component SC, with each component defined on one of the three CFPs around the $M_2$ points of the hexagonal Brillouin zone of the CDW state. The inter-pocket coupling of the Cooper pairs is described by a complex Josephson interaction $Je^{i\varphi}$. The phase of the Josephson coupling $\varphi$ is induced by the LC order. Lower panel: The ground state energy of the superconductor is plotted as a function of the Josephson phase $\varphi$, showing the evolution among roton, isotropic, and anti-roton SC states with different angular momentum $L$ as labeled. Real space configurations of the SC states are shown on the kagome lattice with the onsite arrows denoting the phase of the SC order parameter.
    The arrowed circles indicate the vortices and antivortices forming the V-AV lattice in the roton and anti-roton superconductors.}
	\label{roton_schematic}
\end{figure}

In contrast to a uniform electric current, which is forbidden in the ground state by Bloch's theorem~\cite{1949Bohm}, the staggered LC at nonzero CDW momentum generates staggered loop supercurrent in the SC state. The induced orbital magnetic flux is staggered on the lattice scale and not subject to Meissner effect, which is fundamentally different from an external magnetic field. On physical grounds, the supercurrents can modulate the phase of the SC order parameter and create vortex-antivortex (V-AV) pairs tightly bound on the lattice scale. In the presence of a nonzero center-of-mass momentum pair density wave with a wavevector $Q_{p}$, an ordered V-AV lattice with a lattice constant $2\pi/Q_p$ can emerge with circulating loop supercurrents \cite{2022Zhoub}. Here, in contrast, we investigate how the staggered LC in the CDW normal state determines the nature of the zero-momentum pairing SC state on the kagome lattice.

The pairing instability is studied first for infinitesimal pairing interactions to understand the SC phase structure derived from the LC Chern metal with three CFPs. Fig.~\ref{roton_schematic} shows schematically that pairing on the CFPs realizes a three-component vector superconductor, where the Cooper pairs on different CFPs are coupled by {\em complex} Josephson couplings due to the TRS breaking LC. As a function of the phase of Josephson coupling, the free energy shows that the SC ground state evolves among three different SC phases with different out of plane angular momentum $L$: the $L=\pm2$ chiral states where the phases of the three components are locked at $120^\circ$ with opposite chirality and a $L=0$ state (Fig.~\ref{roton_schematic}). 
We obtain the phase diagram and elucidate the role of the LC-induced discrete quantum geometry governing the Josephson phase, which is associated with the 3-fold rotation generating the sublattice permutation group. 

Self-consistent Bogoliubov-de Gennes (BdG) calculations are then performed at finite pairing interaction strengths. The real space SC order in the $L=\pm2$ states exhibits complex phase windings and loop supercurrents circulating around an emergent V-AV lattice (Fig.~\ref{roton_schematic}). We term this novel SC state a roton superconductor. The concept of a roton, i.e. a tightly-bound V-AV pair, was proposed~\cite{1949Landau,1956Feynman} for a homogeneous superfluid, whose density excitations exhibit a roton minimum at a nonzero momentum corresponding to spatially modulated V-AV pair excitations. The roton superconductor is thus characterized by roton condensation to a V-AV lattice in addition to the condensation of Cooper pairs.

The roton superconductor exhibits intra-unit-cell SC order parameter modulations in real space and anisotropic SC gap in momentum space, visible in the tunneling density of states measured by STM. We show that it is a novel topological superconductor 
supporting chiral edge states carrying nonzero electric currents along domain walls and the sample boundary. 
We discuss the extraordinary SC properties in connection to recent experimental evidence for a TRS breaking chiral SC state
in CsV$_3$Sb$_5$, exhibiting an anisotropic SC gap, pair density modulations, and zero-field SC diode effect.

The internal chiral phases of the three-component roton superconductor can be mapped to the frustrated XY model on the kagome lattice with a fixed chirality/vorticity
due to the normal state LC order, or more precisely to a Josephson junction kagome wire network in an applied frustrating magnetic field. Intriguingly, the supercurrent phase slips generate low-energy fractional $1/3$ vortex and antivortex pair excitations. The proliferation of such internal chiral phase fluctuations gives rise to an extended region of fluctuating charge-2e superconductivity.
We find that charge-6e bound states formed by three Cooper pairs, each residing on a different CFP, decouple from the internal phase fluctuations. The implications for the observed charge-6e flux quantization and SC correlations will be discussed in the fluctuating region of the roton superconductor.

\begin{figure*}
	\includegraphics[width=\textwidth]{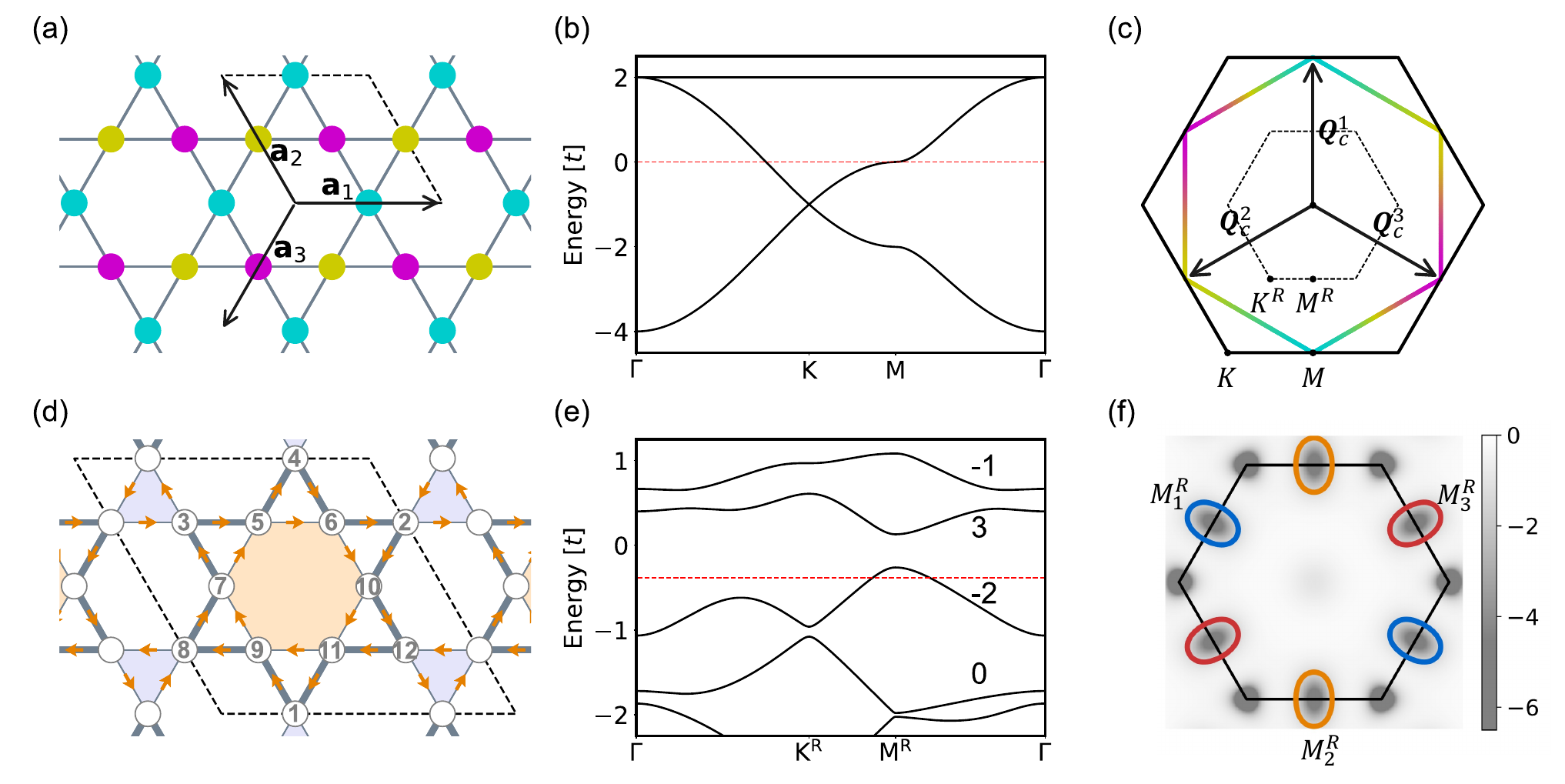}
	\caption{{\bf Kagome lattice at vH filling and $2\times2$ CDW state with LC order}. (a) The $1\times1$ kagome unit cell with three sublattices $\alpha=(1,2,3)$ depicted in (cyan, yellow, pink), respectively. The unit cell vectors $\bm{a}_\gamma$ are plotted as black arrows. (b) Energy dispersion along high-symmetry lines in the single-orbital model. The red dashed line indicates the chemical potential at vH band filling of $n=5/12$. The Fermi level crosses the p-type vH singularity. (c) Fermi surface at vH filling with its sublattice content indicated in colors. Three nesting vectors $\bm{Q}_{c}^{\gamma}$ are labeled by black arrows. (d) The $2\times2$ CDW unit cell (dashed parallelogram) with $12$ sublattice sites labeled by numbers. The thickness of the bond indicates the amplitude of the real part of the complex CDW, and the arrows on the bond indicate the direction of the persistent LC order. (e) Energy dispersion along the high-symmetry lines in reduced Brillouin zone of the CDW state. The topological Chern number is labeled next to each band. The Fermi level (red dashed line) is shifted below the vH filling, which corresponds to doping the Chern insulator into the partially filled Chern band, creating the Chern metal with hole-like CFPs. (f) CFPs and Berry curvature distribution of the LC CDW in the $2\times2$ folded zone. The three CFPs around the $M^{\text{R}}_i$ points are depicted in blue, orange and red, respectively. The dispersion and Berry curvature distributions are obtained with $t_{\text{cdw}}=0.1+0.25i$ and the band filling is $n=4.9/12$.}
	\label{basics}
\end{figure*}

\section{Model for pairing in LC Chern metal}

We study the correlated one-orbital model on the kagome lattice given by the effective Hamiltonian,
\begin{equation}
	H=H_\text{K}+H_{\text{CDW}}+H_{\text{SC}},
    \label{H}
\end{equation}
where $H_\text{K}$ is the tight-binding part, $H_{\rm CDW}$ generates the CDW with LC order, and $H_{\rm SC}$ describes the pairing interactions on the CFPs. 

The tight-binding part is given by
\begin{equation}
	H_\text{K}=-t\sum_{\bm{r},(\alpha\beta\gamma)} (c_{\bm{r}\alpha}^\dagger c_{\bm{r}\beta}+ c_{\bm{r}\alpha}^\dagger c_{\bm{r}-\bm{a}_\gamma\beta} + \text{h.c.})-\mu  \sum_{\bm{r},\alpha}
    c_{\bm{r}\alpha}^\dagger c_{\bm{r}\alpha},
    \label{tb-hopping}
\end{equation}
where $t$ is the nearest neighbor (nn) hopping and $\mu$ is the chemical potential controlling the electron filling.
Here $\bm{r}$ runs over the unit cells, each containing $3$ sublattice sites indexed by $\alpha,\beta\in\{1,2,3\}$ as depicted in Fig.~\ref{basics}(a), together with the lattice vectors $\bm{a}_\gamma$. The spin index is left implicit here. The sublattice indices are summed over in the hopping term according to $(\alpha\beta\gamma)\in\{(123),(231),(312)\}$. The kagome band structure of the one-orbital $H_{\rm K}$ is plotted in Fig.~\ref{basics}(b) along the high-symmetry paths, showing the vH singularities at the $M$ points in the original Brillouin zone (BZ) in Fig.~\ref{basics}(c).

\subsection{Loop-current CDW}
The kagome lattice has a unique property. At $\mu=0$, the
Fermi level crosses the ``pure-type'' (p-type) vH singularity as shown in Fig.~\ref{basics}(b), which are located at the $M$ points touched by the hexagonal Fermi surface (FS) shown in Fig.~\ref{basics}(c) with color-coded sublattice contents. Clearly, the electronic states of the p-type vH singularity are sublattice polarized, i.e. the electrons occupying an $M$ point reside exclusively on one sublattice. This is due to the physics of sublattice interference effects~\cite{2012Thomale}. The nesting between the p-type vH singularities by the wave vectors $\bm{Q}_c^\gamma=\frac12\bm{G}_\gamma$, where $\bm{G}_\gamma$ are the Bragg vectors
shown in Fig.~\ref{basics}(c), thus favors a $2\times2$ bond-ordered CDW state.
The triple-$\bm{Q}$ CDW order can be described by the CDW Hamiltonian~\cite{2022Zhoub},
\begin{equation}
    H_{\text{CDW}}=\sum_{\bm{r}(\alpha\beta\gamma)}t_{\text{cdw}}\cos(\bm{Q}_c^\gamma \cdot \bm{r})(c_{\bm{r}\alpha}^\dagger c_{\bm{r}\beta} - c_{\bm{r}\alpha}^\dagger c_{\bm{r}-\bm{a}_\gamma \beta})+\text{h.c.}
	\label{Hamil_cdw}
\end{equation}
with the $C_6$-symmetric ($C_3$ plus inversion) CDW amplitude $t_{\text{cdw}}$. For real $t_{\text{cdw}}$, $H_{\text{CDW}}$ produces the bond ordered real CDW, exhibiting Star-of-David (SD) modulations for $t_{\text{cdw}}>0$ and inverse-SD (ISD) for $t_{\text{cdw}}<0$. A complex $t_{\text{cdw}}=t_{\text{cdw}}^\prime+it_{\text{cdw}}^{\prime\prime}$ introduces
the staggered LC order, with persistent charge current running on the bonds in the $2\times2$ unit cell~\cite{2022Zhoub} as shown in Fig.~\ref{basics}(d).

The Hamiltonian $H_K+H_{\rm CDW}$ in the LC CDW state can be diagonalized by unitary transformations,
\begin{equation}
	c_{\bm{r}\alpha\sigma}=\sum_{n\bm{k}} e^{i\bm{k}\cdot\bm{r}_\alpha}u_{\alpha n}^{\bm{k}}f_{n\bm{k}\sigma}.
\label{uk}
\end{equation}
Here $\bm{r}_\alpha$ stands for the coordinate of the $\alpha$ sublattice in unit cell $\bm{r}$ and $n$ is the band index. The sum over $\bk$ runs over the reduced BZ and is normalized by the number of $\bk$ points through out the paper. This leads to
\begin{equation}
    H_{\text{K}}+H_{\text{CDW}} = \sum_{n\bm{k}\sigma} E_{n\bm{k}} f_{n\bm{k}\sigma}^\dagger f_{n\bm{k}\sigma},
\end{equation}
where $f_{n\bm{k}\sigma}^\dagger$ creates a spin $\sigma$ quasiparticle in the $n$-th band at momentum $\bm{k}$ with band energy $E_{n\bm{k}}$. The information about the LC is now contained in the complex Bloch wave function $u_{\alpha n}^{\bm{k}}$.

The CDW with LC order breaks TRS and the quasiparticle bands  $E_{n\bm{k}}$ acquire topological Chern numbers calculated from $u_{\alpha n}^{\bm{k}}$ and marked in Fig.~\ref{basics}(e). At vH filling, this gives rise to a CDW Chern insulator. When the Fermi level lies close to but below the vH filling, it crosses the top of the partially filled Chern band in Fig.~\ref{basics}(e), leading to a LC Chern metal with three CFPs~\cite{2022Zhoub} shown
in Fig.~\ref{basics}(f).
The hole-like CFPs are centered around the $M^{\text{R}}_i\,(i=1,2,3)$ points in the $2\times2$ reduced BZ and carry concentrated Berry curvature.
We note that three hole-like Fermi pockets at the same locations have been observed recently by combined ARPES and STM quasiparticle interference imaging~\cite{2023Ilija-pocket}.

\subsection{Pairing Interactions}

We consider the effective pairing Hamiltonian $H_{\rm SC}$ in Eq.~\eqref{H} given by,
\begin{equation}
    H_{\text{SC}}=-W_m\sum_{\langle \bm{r}\alpha,\bm{r}^\prime\beta\rangle_m}\hat n_{\bm{r}\alpha} \hat n_{\bm{r}^\prime\beta},
	\label{Hamil_int}
\end{equation}
where $\hat n_{\bm{r}\alpha}=\sum_\sigma c_{\bm{r}\alpha\sigma}^\dagger c_{\bm{r}\alpha\sigma}$ is the number operator on site $\alpha$ and $W_m>0$ describes attractions between $m$-th nn denoted by $\langle \bm{r}\alpha,\bm{r}^\prime\beta\rangle_m$, with $m=0$ corresponding to onsite attraction. For simplicity, we will use $\langle \alpha\beta\rangle_m$ to denote $\langle \bm{r}\alpha,\bm{r}^\prime\beta\rangle_m$ in the following.

\begin{figure}
    \centering
    \includegraphics[width=\columnwidth]{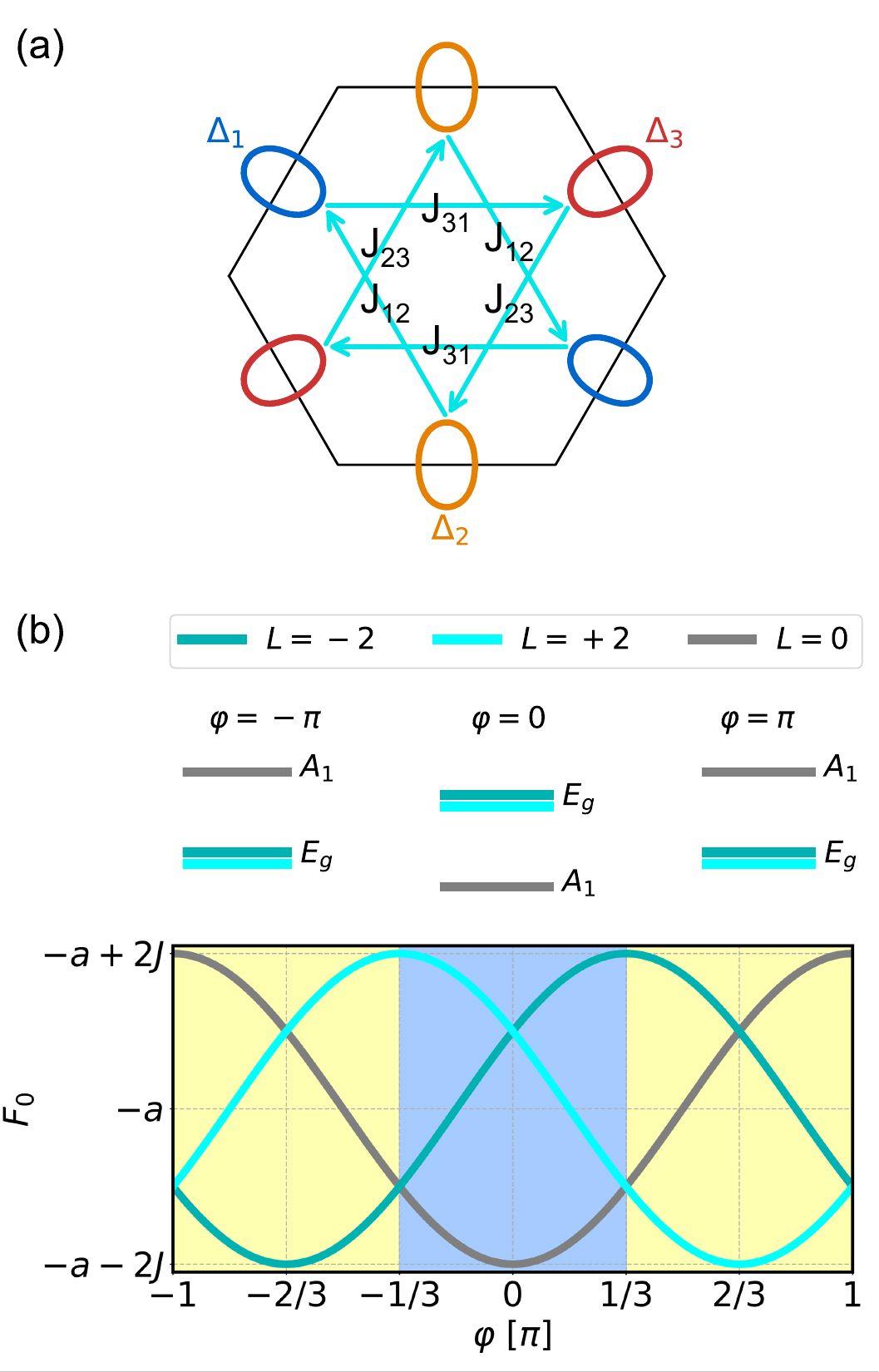}
    \caption{{\bf Pairing on CFPs with Josephson coupling}. (a) Multicomponent superconductor with vector order parameter $(\Delta^f_1, \Delta^f_2, \Delta^f_3)$ defined on the three CFPs. The inter-component complex Josephson coupling is denoted by $J_{ij}$, as shown by the arrows. (b) Free energy derived from the eigenvalues of the $3\times3$ pairing matrix in Eq.~\eqref{33mat} as a function of the Josephson phase $\varphi$. The degeneracy of the two chiral states at $\varphi=0,\pm\pi$ is lifted by the complex Josephson phase. The yellow and blue background fill the areas where the chiral $120^\circ$ state or the $(111)$ state is the ground state, respectively.}
    \label{Josephson_TRS}
\end{figure}

\section{SC instability analysis}
\subsection{Pairing on the CFPs}
As is known from the BCS theory, superconductivity manifests itself as a Fermi surface instability. Thus, we proceed to determine the SC instability of the CFPs by projecting $H_{\rm SC}$ to the pairing channel of the quasiparticles on the CFPs ~\cite{2023varma-wang},
\begin{align}
    H_{\rm SC}=-W_m \sum_{\bm{k},\bm{k}^\prime}\sum_{\langle \alpha\beta\rangle_m}
    \psi_{\alpha\beta}^{\bm{k}\ *}\psi_{\alpha\beta}^{\bm{k}^\prime}\hat \Delta^{f\dagger} (\bm{k}) \hat\Delta^f(\bm{k}^\prime)
    \label{int-proj}
\end{align}
where $\psi_{\alpha\beta}^\bk=\frac{1}{\sqrt{2}}[u_{\alpha n}^{\bm{k}}u_{\beta n}^{-\bm{k}}e^{i\bm{k}\cdot(\bm{r}_{\alpha}-\bm{r}_\beta)}+(\alpha\leftrightarrow\beta)]$
is the wave function of the Cooper pairs created by the spin-singlet pairing operator $\hat \Delta^{f\dagger}(\bm{k})=\frac{1}{2} \epsilon_{\sigma\sigma^\prime}f_{n\bm{k}\sigma}^\dagger f_{n,-\bm{k},\sigma^\prime}^\dagger$.
The band index $n$ corresponds to the partially occupied Chern band, which will be dropped hereafter, and the $\bk$ sum runs over the CFPs. The effects of the TRS breaking LC order and the Berry curvature in the single particle wave function $\vert u^{\bm{k}}\rangle$ are contained in the pair wave function $\vert \psi_m^{\bm{k}}\rangle$ in the
vector space spanned by the $m$-th nn pairs $\{\langle \alpha\beta\rangle_m\}$,
which encodes the important sublattice dependence on the kagome lattice.

Since $\bm{k}$ resides on the three CFPs, labeled by $i=1,2,3$, we have a three-component vector superconductor described by three quasiparticle pairing order parameters $\Delta^f_i(\bk_i)\equiv \langle\hat\Delta^f_i(\bm{k}_i)\rangle$.
To determine the leading SC instability, it is sufficient to keep the quadratic terms in the Ginzburg-Landau free energy,
\begin{eqnarray}
  &&  F_0=-\sum_{\bk_i,\bk_j} \Lambda({\bm{k}_i},
    {\bm{k}_j})\Delta_i^{f*}(\bm{k}_i)
    \Delta_j^f (\bm{k}_j),
    \label{freeE}\\
  && \Lambda({\bm{k}_i},
    {\bm{k}_j})=W_m\sum_{\langle \alpha\beta\rangle_m} \psi_{\alpha\beta}^{\bm{k}_i*} \psi_{\alpha\beta}^{\bm{k}_j}
  =W_m\langle \psi_m^{\bm{k}_i}\vert \psi_m^{\bm{k}_j}\rangle.
   \label{Lambda}
\end{eqnarray}
To make analytical progress, we note that the pairing function can be treated as isotropic approximately due to the smallness of the CFPs, i.e. $\Delta^f_i(\bk_i)\approx\Delta^f_i$, forming a three-component vector pairing order parameter $\bm{\Delta}=(\Delta^f_1,\Delta^f_2,\Delta^f_3)^T$. This is quantitatively justified as the leading order contribution in a partial wave expansion analysis in Appendix~\ref{partial} to account for the intra-pocket pairing structure. The anisotropy will be taken into account in the full numerical calculations.

The sum over the momenta on the CFPs in Eq.~\eqref{freeE} can be carried out to arrive at the element $\Lambda_{ij}=\sum_{\bk_i,\bk_j}\Lambda({\bm{k}_i},
    {\bm{k}_j})$ of the $3\times3$ pairing matrix
    $\bm{\Lambda}$.
The free energy can thus be written as
\begin{equation}
 F_0=-\bm{\Delta}^\dagger \bm{\Lambda}\bm{\Delta}, \quad  \bm{\Lambda}=
    \left(\begin{array}{ccc}
        a & J_{12} & J^*_{31}\\
        J_{12}^* & a & J_{23} \\
        J_{31} & J_{23}^* & a
    \end{array}\right),
    \label{33mat}
\end{equation}
where the diagonal term $a$ in the pairing matrix $\bm{\Lambda}$ describes the intra-pocket pairing of the quasiparticles. As depicted in Fig.~\ref{Josephson_TRS}(a), the off-diagonal term $J_{ij}$ represents the Josephson coupling between the Cooper pairs on two different pockets,
\begin{equation}
    J_{ij}=W_m\sum_{\bk_i,\bk_j}\langle \psi_m^{\bk_i}|\psi_m^{\bk_j}\rangle.
    \label{josephsoncoupling}
\end{equation}
Because of the TRS breaking LC, the Josephson coupling is complex. Under the 3-fold rotation symmetry, we have
$J_{ij} = Je^{i\varphi}$, where $\varphi$ is the Josephson phase. Thus, pairing over the three CFPs is described by the 3-dimensional complex representation of the cubic group, in contrast to the real representation for multicomponent BCS superconductors~\cite{1999Agterberg}.

The pairing matrix $\bm{\Lambda}$ can be diagonalized analytically. Since $\bm{\Lambda}$ commutes with the operator ${\hat C}_3$ for 3-fold rotation for all values of $\varphi$, the eigenstates are those of the 3-fold rotation. This leads to 
three symmetry allowed pairing eigenstates listed below, together with the corresponding pairing free energy,
\begin{alignat*}{3}
  \Psi_0&=\frac{1}{\sqrt{3}}(1,1,1)\bm{\Delta}, \quad
  &E_0&=-a-2J\cos\varphi,  \\
 \Psi_{\pm}&= \frac{1}{\sqrt{3}}(1,\omega_\pm,\omega_\pm^2)\bm{\Delta}, \quad &E_{\pm}&=-a-2J\cos(\varphi\pm\frac{2\pi}{3}).
\end{alignat*}
Here, $\Psi_0$ is the zero angular momentum ($L=0$)  isotropic (111) state, whereas $\Psi_{\pm}$ denotes the two chiral
states with $L=\mp2$ and $\omega_\pm = e^{\pm i2\pi/3}$, such that the relative phases of the three pairing components on the CFPs are locked at $120^{\circ}$ with opposite chirality. Note that the phase of the first component is taken out and absorbed into the overall $U(1)$ SC phase not shown explicitly. 

\subsection{Roton superconductors}
The leading SC instability is determined by the lowest energy eigenstate, i.e. the smallest $E_{0,\pm}$. In Fig.~\ref{Josephson_TRS}(b), the free energies of the three pairing states are plotted as a function of the Josephson phase $\varphi$. For $\varphi \in [-\pi/3, +\pi/3]$, the SC ground state is the $L=0$ isotropic $(111)$ state.
At $\varphi=0$, where the 3D real representation of the cubic group splits into a lowest energy 1D ($A_1$) and a 2D ($E_2$) irreducible representation corresponding to two degenerate excited chiral $d\pm id$ SC states~\cite{1999Agterberg} emerging from the 3-pairing components.
As $\varphi$ becomes nonzero due to the LC order, the degeneracy of the two $120^\circ$ states with $L=\pm2$ is lifted. The nonchiral $(111)$ state with $L=0$ can still break TRS (e.g. an $s+is$ state) due to the LC order, which is encoded in the pair wave function $|\psi_m\rangle$. With increasing $\vert\varphi\vert$, the chiral branch with $\varphi\cdot L>0$ continues to lower its energy while that of the $(111)$ state increases. As a result, for $\vert\varphi\vert > \pi/3$, the LC order drives the ground state to the TRS breaking chiral superconductor $\Psi_\pm$ with the internal phases of the three SC components locked at $120^\circ$ as shown schematically in Fig.~\ref{roton_schematic}, which is dubbed as the roton superconductor.

The wave function of the 3-component roton superconductor is given by
\begin{equation}
\vert\Psi_{\text{roton}}\rangle={\cal N} e^{\sum_{j\bm{k}_j} \omega_{j}{g_{\bm{k}_j}}b_{\bm{k}_j}^{\dagger}}
 \vert\text{vac}\rangle,
    \label{chiral-wf}
\end{equation}
where $b_{\bm{k}_j}^\dagger=f_{\bm{k}_j\uparrow}^\dagger f_{-\bm{k}_j\downarrow}^\dagger$ denotes the creation operator of a Cooper pair on the $j$-th pocket under the pairing function $g_{\bk_j}$,
and $\omega_j$ is the relative phase factor. The overall phase of the superconductor is not written out explicitly. The normalization factor ${\cal N}^{-2}=\Pi_{j\bk_j}(1+|g_{\bk_j}|^2)$. The pairing order parameter for the $f$-quasiparticles can be evaluated using the coherent state wave function in Eq.~\eqref{chiral-wf},
\begin{equation}
    \Delta^f_j(\bm{k}_j)\equiv\langle f_{-\bk_j\downarrow}f_{\bk_j\uparrow}\rangle=\frac{\omega_j g_{\bm{k}_j}}{1+|g_{\bk_j}|^2},
    \label{pairing_function}
\end{equation}
which will be determined, together with the pairing function $g_{\bk_j}$, inside the ordered SC state later. For the roton solution, the internal chiral phase factors $\omega_{j}=e^{i \phi_{j}}$ with $\phi_{j}-\phi_{j+1}=\pm2\pi/3$. The wave function of $(111)$ state corresponds to setting all $\omega_j$ equal in Eq.~\eqref{chiral-wf}.

To determine the leading SC instability of the LC Chern metal, we simply need to calculate the Josephson phase in Eq.~\eqref{josephsoncoupling} 
for a given CDW with LC order. This is straightforward, except for a gauge degree of freedom in diagonalizing the LC CDW Hamiltonian to  obtain the quasiparticle wave function $\vert u^{\bk}\rangle$ in Eq.~\eqref{uk}, which we fix by a symmetric gauge choice to keep rotation symmetry manifest as detailed in Appendix~\ref{gauge}. In Fig.~\ref{instability}, we show the calculated Josephson phase $\varphi$ in the phase space spanned by the real and imaginary CDW amplitudes ($t_{\rm CDW}^\prime$, $t_{\rm CDW}^{\prime\prime}$) and determine the leading SC instability phase diagrams for infinitesimal $m$-th neighbor pairing interactions separately. 

\begin{figure}
    \centering
    \includegraphics[width=\columnwidth]{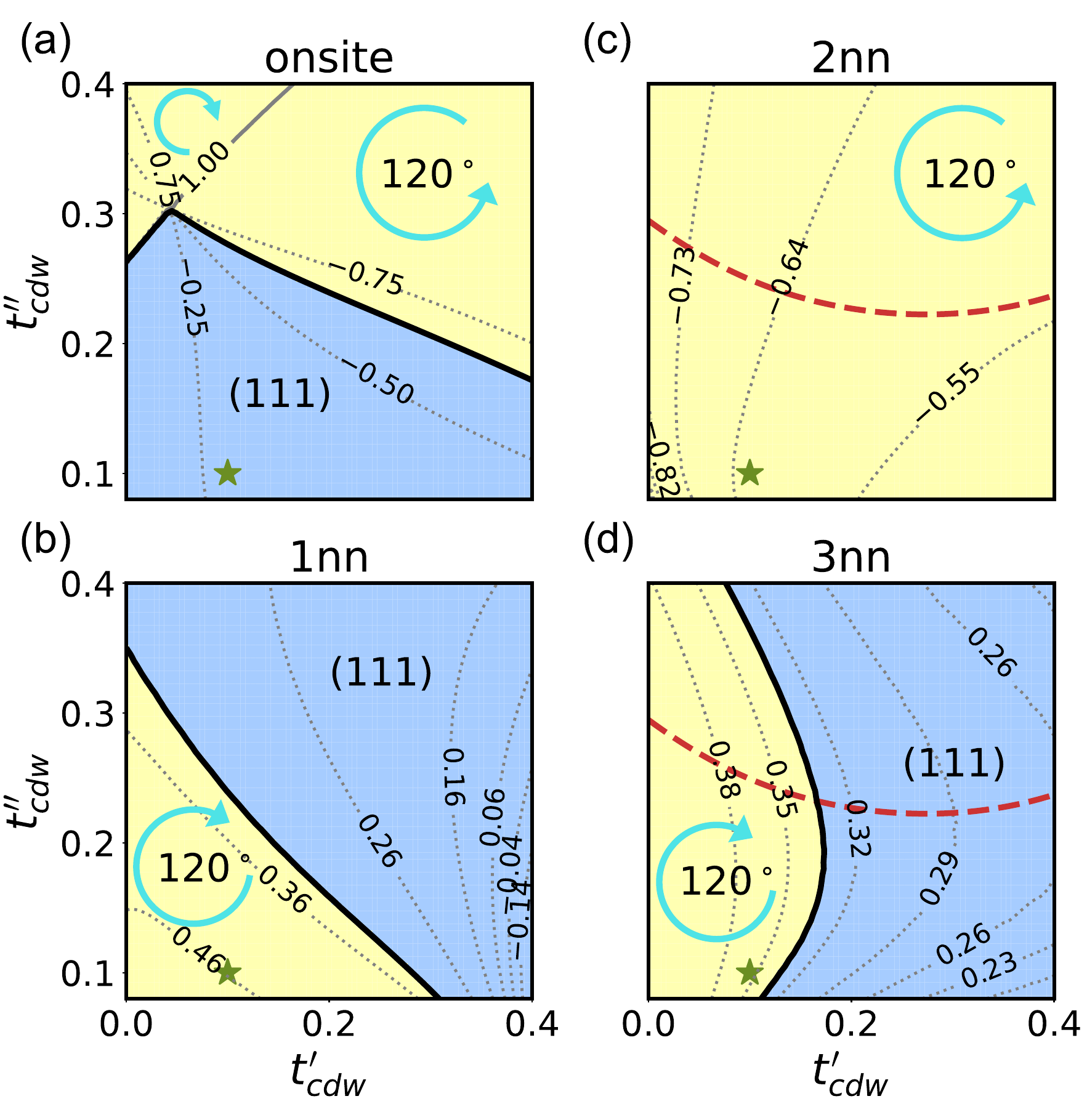}
    \caption{{\bf Pairing instability phase diagram}. Phase diagrams with onsite (a), 1nn (b), 2nn (c) and 3nn (d) attractions, respectively. The ground state pairing configuration, being either chiral $120^\circ$ roton/anti-roton state (yellow) or the isotropic $(111)$ state (blue), is directly obtained from the calculated Josephson coupling phase $\varphi$, whose equipotential contours are shown by the gray dashed lines and labeled in unit of $\pi$. The phase boundaries correspond to $|\varphi|=\pi/3$. The red dashed lines in (c) and (d) indicate the trajectories along which the pocket shape remains the same as in Fig.~\ref{Josephson_TRS}(a). The green star indicates the parameters used in Fig.~\ref{permutation_map}.
    }
    \label{instability}
\end{figure}

For onsite pairing with $m=0$, the isotropic (111) state is favored energetically for small LC, but the $120^\circ$ roton (anti-roton) state becomes the ground state for sufficiently large LC as shown in Fig.~\ref{instability}(a). The calculated Josephson phase is plotted as equipotential contours shown by the gray dashed lines and the phase boundary is determined by $|\varphi|=\pi/3$. Surprisingly, for 1nn, 2nn, and 3nn pairing ($m=1,2,3$), the ground state is already the $120^\circ$ roton (anti-roton) state at small LC as shown in Figs.~\ref{instability}(b-d), which transition to the $(111)$ state for 1nn and 3nn pairing when the real bond CDW amplitude $t_{\text{cdw}}^\prime$ is sufficiently large as shown in Fig.~\ref{instability}(b) and Fig.~\ref{instability}(d). Remarkably, for 2nn pairing in Fig.~\ref{instability}(c), the chiral $120^\circ$ anti-roton state is always found to be the ground state from small to large LC. 

A remarkable feature of the 120$^\circ$ state in Fig.~\ref{instability} is that its chirality or the pair angular momentum $L$ is not uniquely determined by the TRS breaking LC direction in the normal state, but can vary with the range of the pairing interactions and the CDW parameters.
For example, under the fixed LC order, the 2nn pairing produces an anti-roton state in Fig.~\ref{instability}(c) with opposite chirality compared to that of the roton states induced by 1nn and 3nn pairing shown in Fig.~\ref{instability}(b) and (d), respectively.
These results indicate that the sublattice structure and the geometry of the kagome lattice play an important role.
We note that the shape of the Fermi pocket can change considerably for different $t_{\text{cdw}}$. The red dashed line in Fig.~\ref{instability}(c-d) indicates the trajectory along which the CFP maintains its shape as depicted in Fig.~\ref{Josephson_TRS}(a).

\subsection{Role of quantum geometry}

The emergence of the chiral $120^\circ$ state, i.e. the roton superconductor at small LC for 1nn, 2nn, and 3nn pairing indicates that the phase of the Josephson coupling between Cooper pairs on different CFPs can acquire a large value ($|\varphi|>\pi/3$) even when the TRS breaking LC order is relatively weak. The large Josephson phase has a topological origin related to the Berry curvature of the partially filled Chern band hosting the CFPs that couples to the angular momentum of the Cooper pairs.

This can be understood by first considering the simplified case of pairing along a single circular Fermi surface described by,
\begin{equation} H_\circ=\sum_{\bm{k}\bm{k}^\prime}V_{\bm{k}\bm{k}^\prime} \hat{\Delta}^\dagger({\bm{k}})\hat{\Delta}({\bm{k}^\prime}),
\end{equation}
where $\bm{k}$ and $\bm{k}^\prime$ run around the Fermi circle. We have demonstrated in Eq.~\eqref{freeE} that the pairing interaction can be expressed as an inner product of the two-body Bloch wave function, i.e. $V_{\bm{k}\bm{k}^\prime}=\langle\psi^{\bm{k}}|\psi^{\bm{k}^\prime}\rangle$. For two infinitesimally close Cooper pairs,
$\bm{k}^\prime=\bm{k}+d\bm{k}$ or equivalently in terms of the angular coordinate associated with rotation $\theta_{\bm{k}^\prime}=\theta_{\bm{k}}+d\theta$, the pairing interaction can be expanded using the angular momentum operator $\hat{L}_z=i\partial_\theta$ :
\begin{align}
    V_{\theta,\theta+d\theta}&=
     \langle\psi(\theta)|\psi(\theta+d\theta)\rangle \notag\\
     &=1-i\langle\psi(\theta)|\hat{L}_z|\psi(\theta)\rangle d\theta+{\cal O} (d\theta^2)  \notag\\
     &=\langle\psi(\theta)|e^{-i\hat{L}_zd\theta}|\psi(\theta)\rangle.
     \label{V-continuous}
\end{align}
As a result, $V_{\theta,\theta+d\theta}=\exp(-iA_\theta d\theta)$, where $A_\theta=i\langle\psi(\theta)| \partial_\theta | \psi(\theta)\rangle$ is the Berry connection along the Fermi circle. Thus, the phase of the Josephson coupling between the two Cooper pairs is precisely the rotation Berry phase $\gamma=A_\theta d\theta$ accumulated along the infinitesimal segment of the Fermi circle.

The Josephson coupling between the CFPs in Eq.~\eqref{33mat} can be described in a similar fashion, except that the CFPs are related by a discrete 3-fold rotation of $\bk$, instead of a continuous translation (rotation). Specifically, the Josephson phase, written in terms of the inner product of the pair wave functions, can be expressed in terms of the angular momentum operator $\hat{L}_z$,
\begin{align}
 \langle \psi_m^{\bk_i}|\psi_m^{\bk_j}\rangle=  \langle \psi_m^{\bk_i}|e^{-i{2\pi\over3}\hat{L}_z}|\psi_m^{\bk_i^\prime}\rangle,
 \label{rotation_phase}
\end{align}
where $e^{-i{2\pi\over3}\hat{L}_z}$ generates a rotation of $2\pi/3$ in $\bk$-space that connects $\bk_i^\prime$ on the $i$-th pocket to $\bk_j$ on the $j$-th pocket. As detailed in Appendix~\ref{permutation}, the discrete $\bk$-space rotation generated by $e^{-i{2\pi\over3}\hat{L}_z}$ is related to the sublattice permutation generated by the operator $\hat{P}_3$, because the full 3-fold rotation $\hat{R}_3=\hat{P}_3\otimes e^{i\hat{L}_z2\pi/3}$ is a symmetry of the LC-CDW Hamiltonian. The sublattice permutation operator $\hat{P}_3$ acts on the sublattice components of the vector $|\psi_m^\bk\rangle$ by permuting them to form the permutation group within discrete 3-fold rotations around the center of the SD shown in Fig.~\ref{basics}(d) in real space. 
Specifically, $\hat{P}_3|\psi_m^\bk\rangle=|\psi_{[m]}^\bk\rangle$, where the subscript $[m]$ denotes the group element of the vector basis $\{\langle\alpha\beta\rangle_{[m]}\}$ corresponding to permuting that of $\{\langle\alpha\beta\rangle_m\}$ under the 3-fold rotation.
The Josephson coupling in Eq.~\eqref{josephsoncoupling} can thus be written as:
\begin{equation}
 J_{ij}=W_m\sum_{\bk_i,\bk_j}\langle \psi_m^{\bk_i}|\psi_m^{\bk_j}\rangle =W_m\sum_{\bk_i,\bk_i^\prime}\langle  \psi_m^{\bk_i}|\hat P_3| \psi_m^{\bk_i^\prime}\rangle,
\label{j-berry}
\end{equation}
which is determined by the vector overlap of the pair wave function under one permutation operation. 

\begin{figure}
    \centering
    \includegraphics[width=\columnwidth]{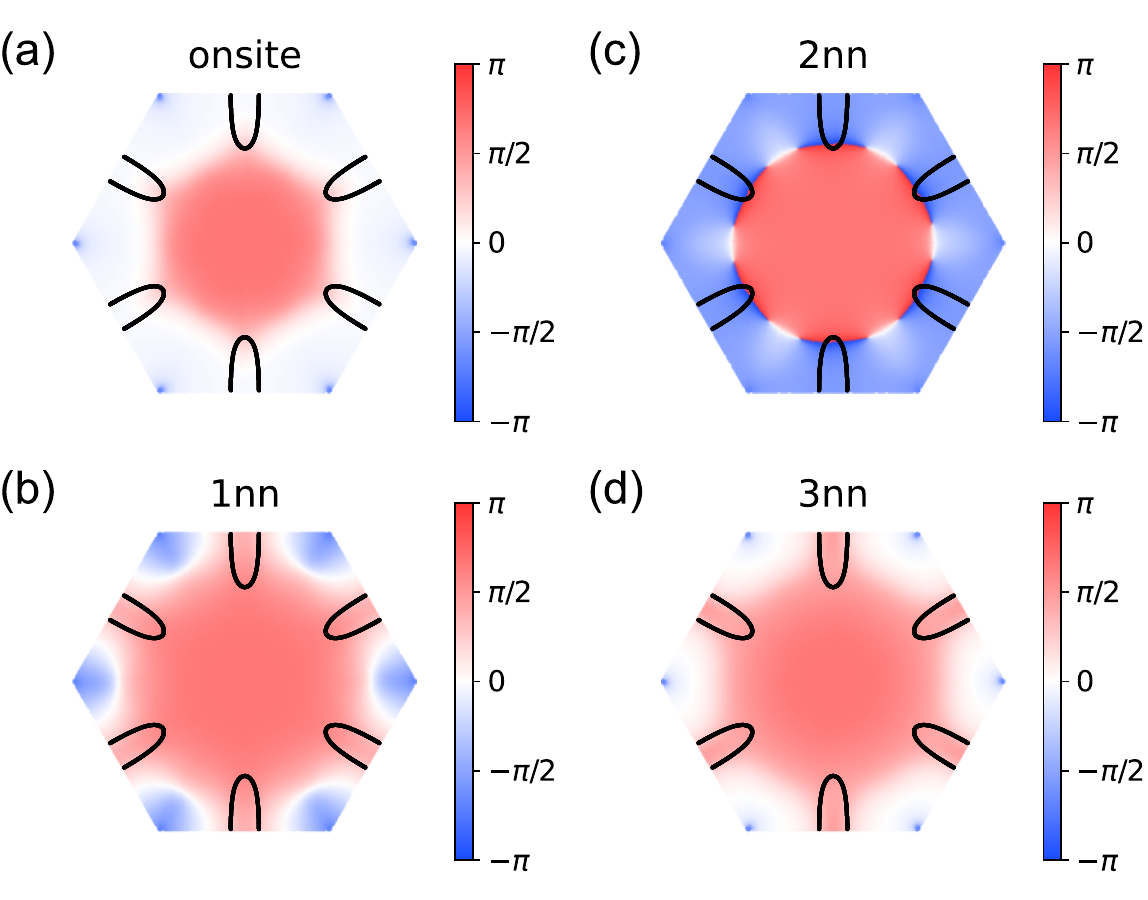}
    \caption{{\bf Berry phase of sublattice permutation}. Map of the sublattice permutation phase $\phi_P(\bk)=\arg (\langle  \psi_m^{\bk}|\hat P_3| \psi_m^{\bk}\rangle)$. The CDW parameters $t_{\text{cdw}}=0.1+0.1i$ are located at the red stars in the phase diagrams in Fig.~\ref{instability}. (a-d) show the permutation phase map for onsite, 1nn, 2nn and 3nn pair wave function $|\psi^{\bm{k}}_m\rangle$, respectively. The Fermi pockets are superimposed in black solid lines.}
    \label{permutation_map}
\end{figure}

Intriguingly, Eq.~\eqref{j-berry} shows that the Josephson coupling is expressed as a double-integral over $\bk_i$ and $\bk_i^\prime$ on a \emph{single} CFP. 
The contribution from integrating around the CFP along the path $\bk_i=\bk_i^\prime\equiv \bk$ corresponds to the geometrical Berry phase contribution associated with the discrete angular rotation or equivalently that of the sublattice permutation. Defining the permutation phase $\phi_P(\bk)=\arg (\langle  \psi_m^{\bk}|\hat P_3| \psi_m^{\bk}\rangle)$, we find that its integral over the CFP dominates the Josephson phase $\varphi$, since the contributions from the sum over $\bk_i\neq\bk_i^\prime$ are suppressed due to destructive interference and can be neglected to a good approximation. As a result, the topological property of the CFPs in the partially filled Chern band is responsible for the large Josephson phase $\varphi$, highlighting the role of discrete quantum geometry associated with the sublattice permutation group. The Josephson phase $\varphi$ can now be read off directly from the map of the calculated sublattice permutation phase $\phi_P(\bk)$ shown in the reduced BZ in Fig.~\ref{permutation_map} for different near neighbor pairings.

For onsite pairing, the CFPs mainly resides near $\phi_P=0$ in Fig.~\ref{permutation_map}(a), leading to the small Josephson phase and the isotropic (111) SC state at small LC in Fig.~\ref{instability}(a).
On the other hand, for 1nn and 3nn pairing, the pockets are located primarily in the region of large positive permutation phases, generating the large Josephson phase necessary for the chiral roton SC state in response to the small LC shown in Fig.~\ref{instability}(b) and (d). Interestingly, for 2nn pairing, Fig.~\ref{permutation_map} shows that the permutation phase around the CFPs is mostly large and {\em negative}, which explains the opposite chirality of the resulting 120$^\circ$ roton state in Fig.~\ref{instability}(c) compared to the other cases.
These results highlight the important role of the Berry curvature induced by LC order and that of the kagome sublattices in determining the property of the SC states. 

\begin{figure}
    \centering
    \includegraphics[width=\columnwidth]{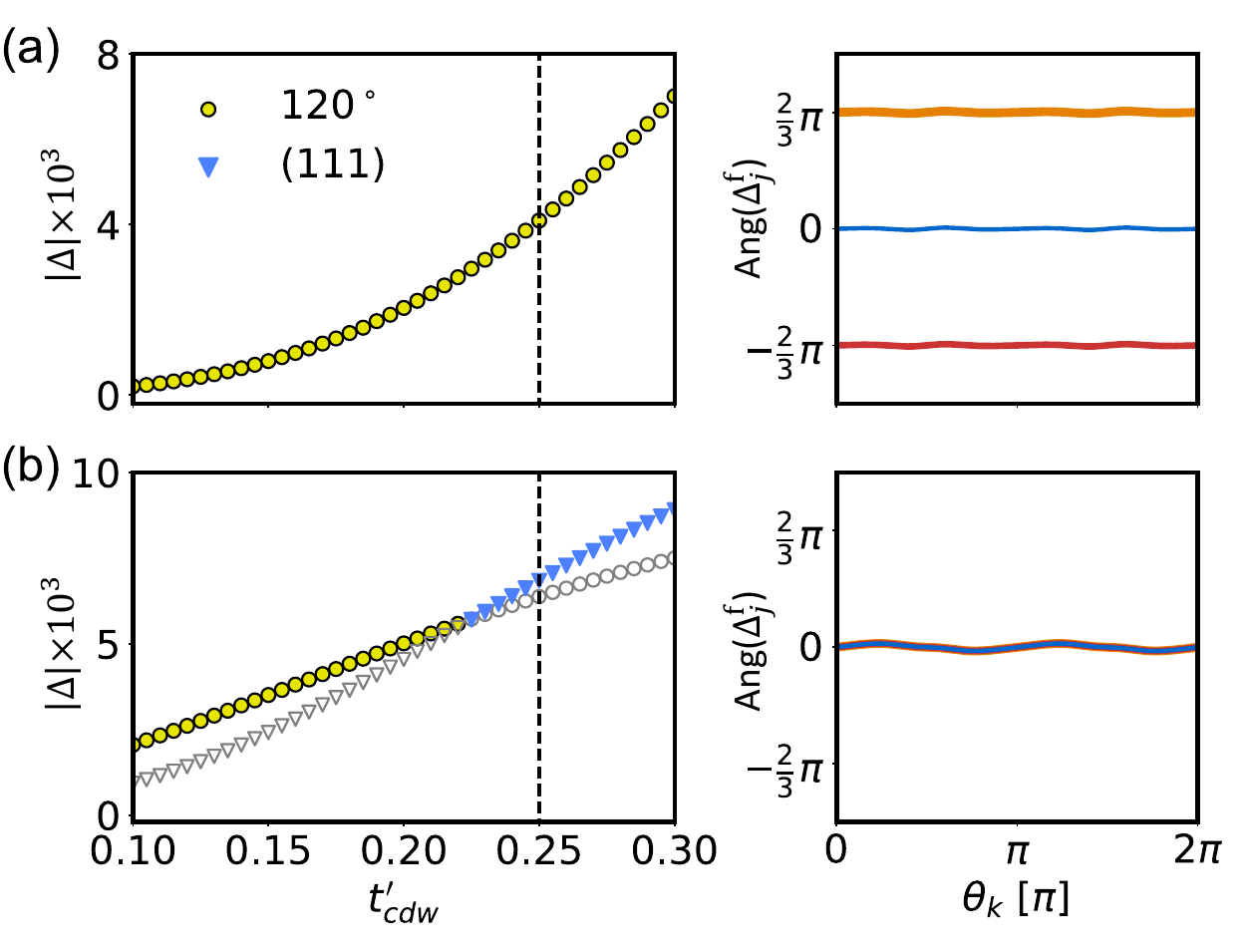}
    \caption{{\bf SC order parameters determined by solving BdG equations}. Self-consistent pairing solutions under $W_2=-1.0$ (a) and $W_3=-1.0$ (b) along the red dashed lines in the phase diagrams shown in Fig.~\ref{instability} (c) and (d). Right panel: the phase of $\Delta^f_j(\theta_{\bk_j})$ along the three CFPs obtained at $t^\prime_{\text{cdw}}=0.25$, as indicated by the dashed lines in the left panel. The pairing phase is plotted as a function of the polar angle $\theta_\bk$ around the corresponding $M_j^\text{R}$ point. The colors are the same as in Fig.~\ref{Josephson_TRS} (a). For 2nn pairing, the anti-roton SC state is reflected by the $120^\circ$ phase differences among the CFPs (top panel), while for 3nn pairing all three CFPs are in-phase in the ground state, in agreement with the $(111)$ SC state (bottom panel).}
    \label{meanfield}
\end{figure}

\section{Properties of charge-2$e$ roton superconductors}

\begin{figure*}
    \centering
    \includegraphics[width=\textwidth]{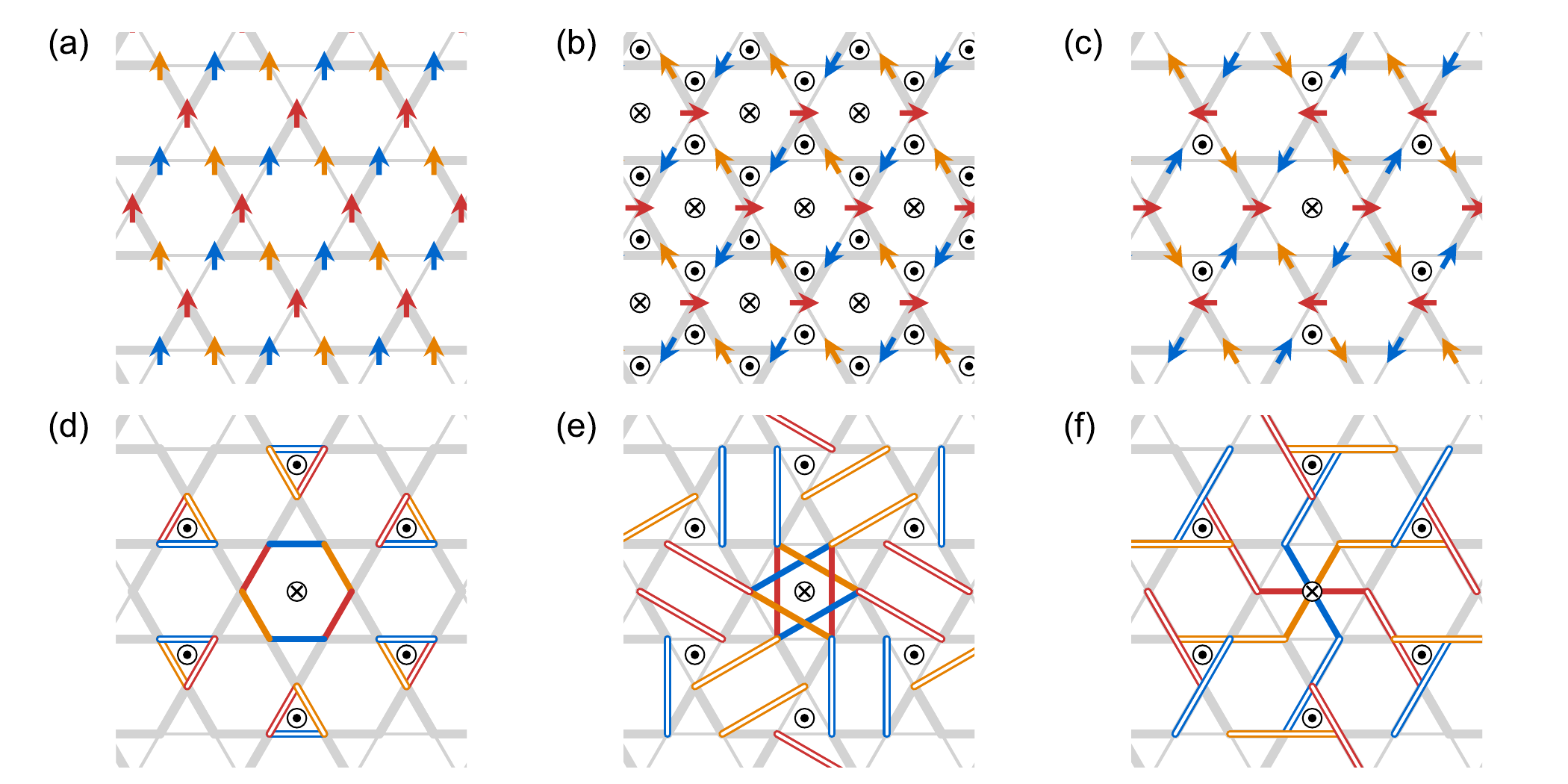}
    \caption{{\bf Order parameter phase distributions and emergent V-AV lattice.} (a-c) Arrows indicate the phase of the onsite pairing order parameter $\Delta({\bf r}_\alpha)$. (a) Isotropic $(111)$ SC state. The sublattice components are locked in-phase. (b)  Roton superconductor with $1\times1$ modulations. The internal phases of the three sublattice components are locked at $120^\circ$.  (c) Roton superconductor with $2\times2$ modulations. The internal phases of the pairing order parameter follow (b) with an additional $\pi$-phase modulation along the three lattice directions. The winding of the SC phases in (b) and (c) gives rise to an emergent hexagonal V-AV lattice. The center of  the vortices are marked by $\odot$ with vorticity $v=+1$ and double antivortices by $\otimes$ with vorticity $v=-2$.
    (d-f) Internal phase distribution of the 1nn, 2nn and 3nn pairing order parameters in the roton superconductor with $2\times2$ modulations, respectively. Solid lines stand for pairing bonds around the double-antivortices and double-solid lines for those around the vortices. The colors (red, blue, yellow) correspond to the internal phases of the pairing order parameter on the bonds ($0$,  $-2\pi/3$,  $2\pi/3$). 
     (b-f) are obtained from the self-consistent solution under $t_{\text{cdw}}=0.1+0.3i$ with $W_m=\{-1,-1,-1,-1\}$. The thickness of the underlying kagome lattice bond in (a-f) represents the real part of the $2\times2$ bond ordered CDW in the normal states.}
    \label{vortex_nn}
\end{figure*}

We next go beyond the instability analysis and study the SC ground state at finite strengths of the pairing interactions $W_m$. The calculations are done directly in real space by solving the BdG equations for the $m$-th nn SC order parameter
\begin{equation}
    \Delta_{\langle \alpha\beta\rangle_m}=\frac12\left[\langle c_{\bm{r}\alpha\downarrow} c_{\bm{r}'\beta\uparrow}\rangle -\langle c_{\bm{r}\alpha\uparrow} c_{\bm{r}'\beta\downarrow}\rangle\right],
    \label{po}
\end{equation}
fully self-consistently as detailed in Appendix~\ref{self-consistent}. As examples, in the left panel of Fig.~\ref{meanfield}, we plot the maximum amplitudes of the 2nn and 3nn pairing bonds ($\max (\vert\Delta_{\langle \alpha\beta\rangle_m}\vert)$) under only $W_2$ or $W_3$, respectively, calculated along the pocket-shape-preserving trajectories indicated by the red dashed lines in 
Fig.~\ref{instability}(c) and (d). In the right panel, the phases of the corresponding quasiparticle pairing order parameters along the CFPs, i.e. the phases of $\Delta^f_j(\bk)$ in Eq.~\eqref{pairing_function}, are plotted. The derivation of $\Delta_j^f(\bk)$ from $\Delta_{\langle\alpha\beta\rangle_m}$ can be found in Appendix~\ref{self-consistent}. Under $W_2$, the solution is always the $120^\circ$ chiral anti-roton state, while there is a phase transition from the $120^\circ$ roton superconductor to the $(111)$ state under $W_3$. These results are fully consistent with the predictions of the instability phase diagram in Fig.~\ref{instability}(c-d).

\subsection{Pair density modulations}
We now turn to the spatial configuration of the self-consistently determined SC order parameter in the chiral roton states emerging from the CDW with LC order specified by $t_{\rm cdw}=0.1+0.25i$. For generality, we consider a specific choice of $W_m$'s where all the $m$-th nn pairing interactions are non-zero, namely $W_m=\{-1,-1,-1,-1\}$. We find that, generic of the chiral roton state, both the phase and the amplitude of the pairing order parameter $\Delta_{\langle \alpha\beta\rangle_m}$ exhibit spatial modulations with distinct $1\times1$ and $2\times2$ periodicity, which are intra-unit-cell SC modulations that do not break lattice translation symmetry of the CDW state. The origin can be traced back to pairing on the CFPs located at the CDW zone boundary, as depicted in Figs.~\ref{roton_schematic}, \ref{basics}(f), and \ref{Josephson_TRS}(a), that involves the reciprocal lattice vector of the CDW lattice (${\bf Q}_c={1\over2}{\bf G}$) and the original kagome lattice (${\bf G}$). We note that 1nn pairing in the single-orbital kagome model with intersite Coulomb interactions has been studied at the $p$-type vHS recently and found to exhibit $2a_0\times2a_0$ chiral SC modulations~\cite{2024Wangqh}. Indeed, both the $2\times2$ and $1\times1$ SC modulations have been observed in the SC gap modulations in KV$_3$Sb$_5$ recently~\cite{2024Yin,2025ligeng}. We refer to these as pair density modulations \cite{2025Wangc,2025Nadj-Perge} since they are SC modulations inside, i.e. intra CDW unit-cell.
They are, however, different from the primary pair density waves observed in CsV$_3$Sb$_5$ with ${4\over3}\times{4\over3}$ periodicity~\cite{2021Chen,2022Zhoub} and in the emergent SC state on the $2\times2$ Cs-reconstructed surface with $4\times4$ periodicity~\cite{chenhui-surface}, both breaking the lattice translation symmetry beyond the $2\times2$ CDW order. 

For simplicity, consider the spin-singlet onsite pairing order parameter corresponding to $m=0$ in Eq.~\eqref{po}, which can be evaluated using the roton wave function in Eq.~\eqref{chiral-wf}. Dropping the spin indices, the singlet pairing order parameter located at $\bm{r}_\alpha$ can be evaluated as:
\begin{align}
\Delta({\bf r}_\alpha)&=\langle\Psi_{\rm roton}\vert c_{\bm{r}\alpha} c_{\bm{r}\alpha}\vert\Psi_{\rm roton}\rangle =
\sum_{j\bk_j}\psi_{\alpha\alpha}^{{\bk_j}}\langle f_{\bk_j} f_{-\bk_j}\rangle 
\nonumber\\
& =\sum_{j\bk_j}\psi_{\alpha\alpha}^{{\bk_j}}\Delta_j^f(\bk_j)=\sum_{j\bk_j}e^{i\phi_j} \psi_{\alpha\alpha}^{{\bk_j}}\frac{g_{\bk_j}}{1+|g_{\bk_j}|^2},
\label{SC-2e}
\end{align}
where the pair wave function $\psi_{\alpha\alpha}^{{\bk_j}}$ was defined in Eq.~\eqref{int-proj}. Clearly, all three pairing components contribute to the local SC order parameter. We show explicitly in Appendix~\ref{connection} that the internal phases $e^{i\phi_j}$ in Eq. ~(\ref{SC-2e}) yield a chiral phase relation among $\Delta(\bm{r}_\alpha)$, i.e. $\Delta(\bm{r}_1):\Delta(\bm{r}_2):\Delta(\bm{r}_3)=e^{i\phi_1}:e^{i\phi_2}:e^{i\phi_3}$ with $\bm{r}_{1,2,3}$, defined in Fig.~\ref{basics}(d), exemplifying three sublattices related by 3-fold rotation.
Indeed, the self-consistently obtained $\Delta({\bm r}_\alpha)$ can be written, up to an overall phase, as a superposition of triple-$\bm{Q}$ modulations along the three crystal lattice directions
\begin{align}
    \Delta (\bm{r}_\alpha)=\sum_{\gamma=1}^3 e^{i\varphi_\gamma} \bigl[ &A_1 \cos \bm{G}_\gamma\cdot(\bm{r}_\alpha-\bm{r}_0)
    \notag\\
    &+A_2 \cos\bm{Q}_c^\gamma\cdot(\bm{r}_\alpha-\bm{r}_0+\delta_\gamma)\bigr].
\label{op}
\end{align}
In the above expression, $\bm{r}_0$ is the center of the triple-$\bm{Q}$ pair density modulations located at the center of the SD, $\varphi_\gamma=(1,e^{\mp i{2\pi}/3},e^{\mp i4\pi/3})$ are the relative chiral phases with $\mp$ for the $L=\pm2$ roton states, $A_{1,2}>0$ denote the magnitudes associated with the $1\times1$ and the $2\times2$ pair density modulations with a phase shift $\delta_\gamma=(\pi,\pi,\pi)$. We derive explicitly in Appendix~\ref{connection} that the ratio among the local pairing order parameters $\Delta(\bm{r}_\alpha)$ related by 3-fold rotation is directly related to $\varphi_\gamma$. For example, consider the three sublattices $(1,2,3)$ as defined in Fig.~\ref{basics}(d), the ratio $\Delta(\bm{r}_1):\Delta(\bm{r}_2):\Delta(\bm{r}_3)=e^{i\varphi_1}:e^{i\varphi_2}:e^{i\varphi_3}$, which is in the same form as written in terms of $\phi_j$ discussed above. We therefore conclude that the relative phase $\varphi_\gamma$ in the triple-$\bm{Q}$ pair density modulations is physically equivalent to the corresponding internal phase $\phi_j$ of the Cooper pairs on the CFPs.

\subsection{Vortex-antivortex Lattice}

In Fig.~\ref{vortex_nn}(a-c), we plot the phase of the onsite SC order parameter $\Delta(\bm{r}_\alpha)$ as the direction of arrows. The three different colored (red, blue, yellow) arrows indicate the SC phase on the three sublattices. The case depicted in Fig.~\ref{vortex_nn}(a) is the non-chiral $(111)$ phase, where the three pairing components are in phase. Intriguingly, as depicted in Fig.~\ref{vortex_nn}(b), the 3-component roton superconductor with $L=+2$ can be understood as three phase uniform pairing states on each of the three sublattices, whose relative phases are locked at $120^\circ$. When $A_2>A_1$, the $2\times2$ part dominates and the local pairing phase on each sublattice gains an additional $\pi$-phase modulation along the three lattice directions as shown in Fig.~\ref{vortex_nn}(c). 

The sublattice correspondence of the three-component roton superconductor is remarkable. Notice that the SC phase winds by $2\pi$ around a triangle, giving rise to a vortex with vorticity $v=+1$ as indicated by $\odot$ in Fig.~\ref{vortex_nn}(b), whereas it winds by $-4\pi$ around a hexagon, giving rise to a double-antivortex with vorticity $v=-2$ as depicted by $\otimes$ . Thus, the phase modulations are described by an emergent $1\times1$ V-AV lattice on the atomic scale. Similarly, the phase modulations in Fig.~\ref{vortex_nn} (c) correspond to an emergent $2\times2$ V-AV lattice delineating the ISD/tri-hexagonal CDW pattern. Since a roton is a tightly bound V-AV pair, the V-AV lattice can be thought as a roton condensate at nonzero momentum ${\bf Q}_{\rm roton}=\{{\bf Q}_c, {\bf G}\}$, and hence the name roton superconductor.
With persistent loop supercurrents circulating around the V-AV lattices, this extraordinary SC state exhibits TRS breaking pure pair density modulations 
at the reciprocal lattice wave vectors. It will be shown below that the roton superconductor is a chiral topological superconductor with edge states carrying electric current.

The V-AV lattice of the roton superconductor is also supported by the intersite pairing order parameters on the $m$-th ($m=1,2,3$) nearest neighbor bonds. The $2\times2$ V-AV lattices are shown in Fig.~\ref{vortex_nn}(d-f), where the pairing order parameters on the $m$-th nn bonds are plotted using solid lines color-coded with their relative phases in red ($0$), blue ($-2\pi/3$), and yellow ($2\pi/3$). For all values of $m$, the central hexagon hosts a double antivortex ($\otimes$) around which the relative phases of the pairing bonds wind by $-4\pi$. The outer triangles of the ISD/tri-hexagonal pattern, on the other hand, host a vortex ($\odot$) around which the relative pairing phases of the bonds denoted by the double solid lines wind by $2\pi$. The spontaneously nucleated V-AV lattice and the circulating staggered loop supercurrents intertwined with the chiral pair density modulations are essential characteristics of the roton superconductor.

\begin{figure}
    \centering
    \includegraphics[width=\columnwidth]{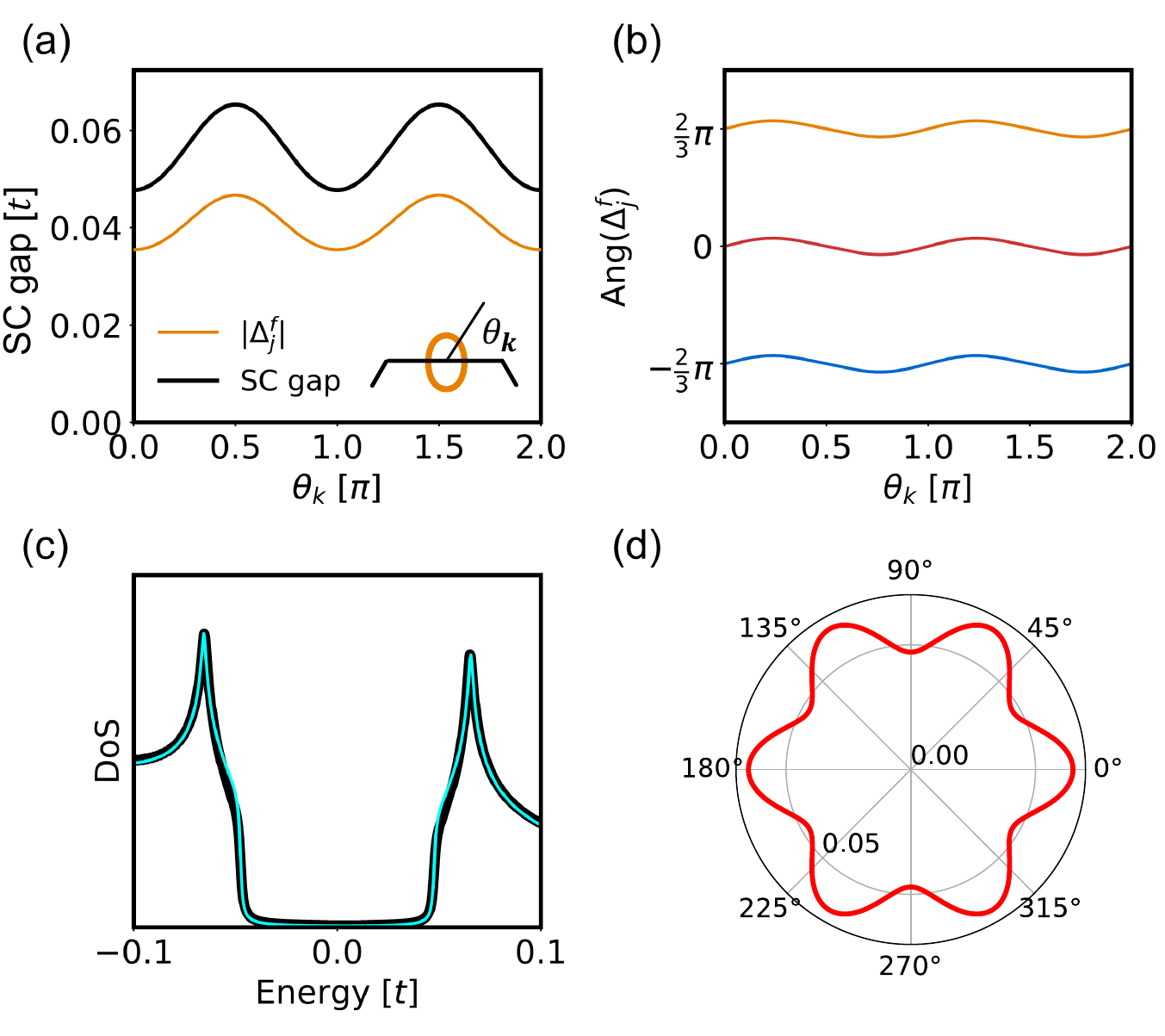}
    \caption{{\bf Anisotropic SC gap and tunneling DoS.} (a) The quasiparticle pairing order parameter magnitude $|\Delta^f_j(\theta_{\bk_j})|$ (orange line) and the SC gap size (black line) along the CFP as a function of the internal angle $\theta_\bk$ illustrated in the inset of (a). (b) The phases of $\Delta^f_j(\theta_{\bk_j})$ along the three CFPs, which are locked at $120^\circ$ in a chiral relation. The colors are the same as in Fig.~\ref{Josephson_TRS} (a). (c) The tunneling DoS in the SC phase obtained from the self-consistent BdG solutions (black line). The cyan line denotes the fitting by the Dynes function with an anisotropic SC gap whose angular dependence is shown in (d). The results are obtained under $t_{\text{cdw}}=0.1+0.3i$ and $W_m=\{-1, -1, -1, -1\}$.
    }
    \label{dos}
\end{figure}

\subsection{Tunneling density of states}

In the ground state, the CFPs are fully gapped in the roton superconductor. However, the SC gap magnitude is found to be anisotropic around each CFP. As an example, we calculate the complex pairing order parameters on the three CFPs in Eq.~\eqref{pairing_function} by solving the BdG equation for pairing interactions $W_m=\{-1,-1,-1,-1\}$. The anisotropic order parameter amplitude around the CFPs is plotted in Fig.~\ref{dos}(a), while the relative phase is locked at 120$^\circ$ as shown in Fig.~\ref{dos}(b). From the BdG quasiparticle dispersion, we extract the SC gap magnitude at the Fermi wave vectors of the CFP, i.e. at the minimum gap locus, which is plotted in Fig.~\ref{dos}(a).
The SC energy gap shows an approximate 30\% gap anisotropy, 
which can in principle be measured by angle-resovled photoemission. 

The SC gap anisotropy can also be measured by STM directly in real space.
To this end, we calculate the local tunneling density of states (DoS),
\begin{equation}
    \rho(\bm{r}_\alpha,\omega) = -{1\over\pi}\sum_\bk\rm{Im}[G_{\alpha\alpha}(\bk,\omega)],
\end{equation}
where $G_{\alpha\alpha}(t)=-i\theta(t)\sum_\sigma\langle [c_{\bm{r}\alpha\sigma}(t),c^\dagger_{\bm{r}\alpha\sigma}]_+\rangle_T$ is the electron Green's function. The tunneling DoS exhibits modulations inside the $2\times2$ unit cell shown in Fig.~\ref{basics}. Average over the 12 sites in the unit cell, we obtain the tunneling DoS $\rho(\omega)$ plotted in Fig.~\ref{dos}(c). The spectrum is characteristic of an superconductor with an anisotropic SC gap, where a pair of coherence peaks delineate the maximum gap ($\Delta_{\rm max}$) with the in-gap shoulders corresponding to the gap minimum ($\Delta_{\rm min}$) on the CFPs. The latter can be extracted from fitting the spectrum by the Dynes function
$\rho(\omega)=P(\omega)\int_0^{2\pi}d\theta \text{Re}\frac{\omega-i\Gamma}{\sqrt{(\omega-i\Gamma)^2+\Delta^2(\theta)}}$, with an anisotropic gap function $\Delta(\theta)=\Delta_1+\Delta_2 \cos(6\theta)$. The function $P(\omega)$ is a power-law function commonly introduced to account for the normal state DoS. An accurate fit is shown in Fig.~\ref{dos}(d) with $\Delta_1=0.0566t$ and $\Delta_2=0.0090t$, corresponding to $\Delta_{\rm max}=0.0656t$
and $\Delta_{\rm min}=0.0476t$. The broadening in the Dynes function is set to $\Gamma=0.001$, which is also the broadening width used in obtaining the DoS plot. The angular distribution of the SC gap is shown in Fig.~\ref{dos}. It is remarkable that such an anisotropic SC gap angular distribution has been detected in CV$_3$Sb$_5$, coexisting with an isotropic gap possibly due to the Sb $p$-orbital in the multi-gap kagome superconductor~\cite{mode-nc2024,zywang-nc,2024Nowack}. 
We note that the local DoS spectrum measured by STM also contains ungapped states coming from the ungapped Fermi surface sections associated with other $d$-orbitals~\cite{2024Yin} and possibly the effects of roton pair density waves and V-AV lattice~\cite{2022Zhoub}. The understanding and description of the unpaired states are currently elusive and subjects of future research.

\subsection{Topological current-carrying edge states}

\begin{figure}
    \centering
    \includegraphics[width=\columnwidth]{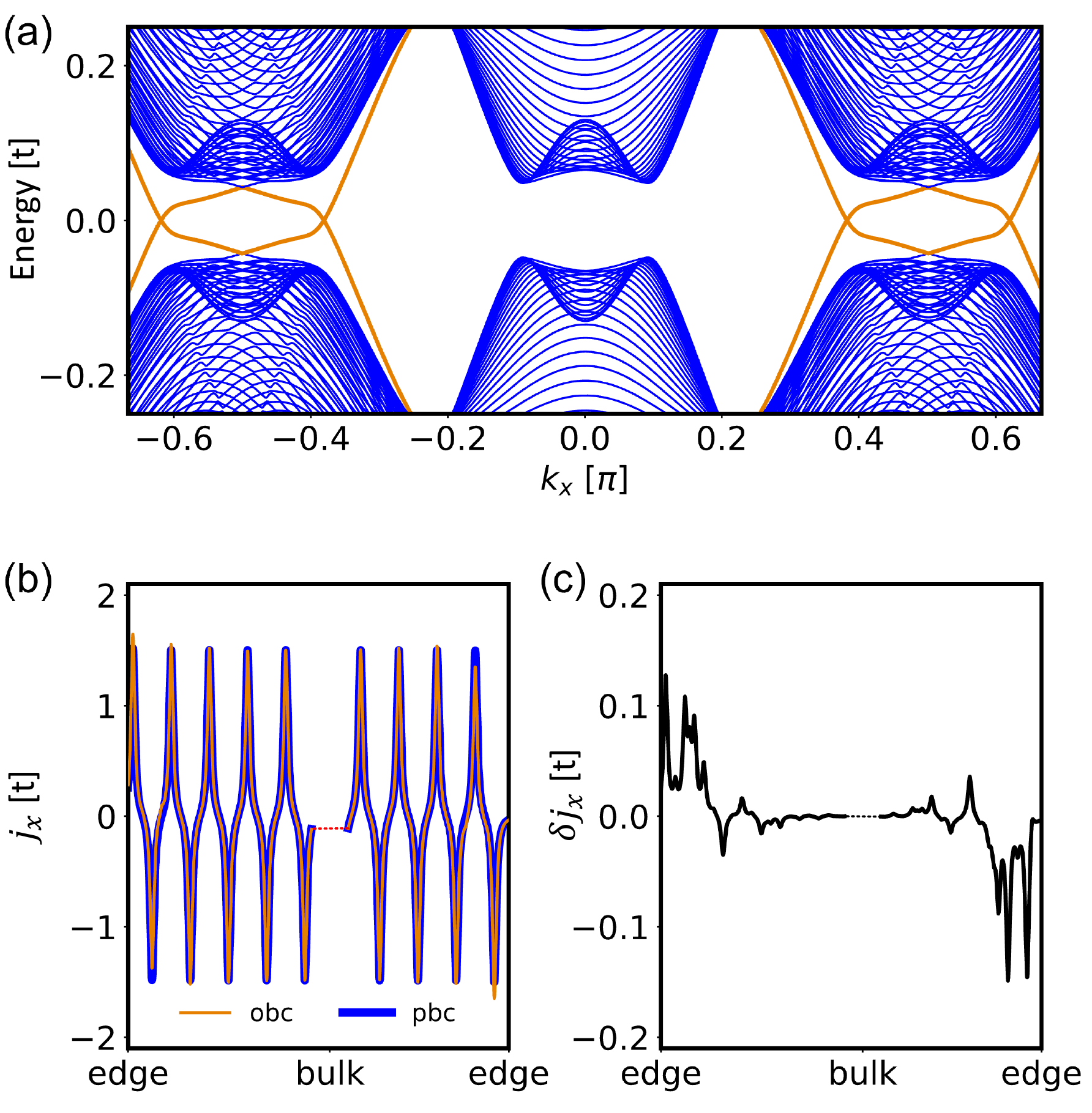}
    \caption{{\bf Chiral edge modes and edge currents of the roton superconductor.} (a) Quasiparticle energy dispersion with $k_x$ solved under open boundary condition in the y-direction and periodic boundary condition in the x-direction. The geometry is the same as the long cylinder used in~\cite{2022Zhoub} with a cylinder structure containing $60$ $2\times2$ unit cells in its y-direction and the number of $k_x$ is set to $800$. The chiral edge modes, plotted in orange, appear inside the SC gap of the bulk states (blue). (b) Current $j_x(y)$ along the $x$-direction as a function of $y$, plotted from one edge of the open cylinder to the other. The current $j_x(y)$ is obtained by calculating the bond current $\bm{j}_{\alpha\beta}$, projecting it to the $x$-direction, and summing over all bonds with the same $y$-location defined at the bond center, i.e. $y=(y_\alpha+y_\beta)/2$. The results for $j_x(y)$ obtained under both periodic (blue curve) and open (orange curve) boundary conditions are shown. Under the open boundary condition, the edge modes carry significant counter propagating charge current as revealed by the difference $\delta j_x(y)=j_{x,\rm obc}(y)-j_{x,\rm pbc}(y)$ between the two boundary conditions shown in (c), concentrating near the two ends of the cylinder. The results are obtained under $t_{cdw}=0.1+0.3i$ and $W_m=\{-1, -1, -1, -1\}$. } 
    \label{current}
\end{figure}

The chiral roton superconductor is topological and can be described by the topological Chern numbers associated with the BdG quasiparticle bands. We find the total Chern number $C=2$ at $t_{cdw}=0.1+0.3i$, which is the same as that of the LC CDW normal state when the hole-like CFPs were fully occupied.
This implies that there should be two sets of topological edge modes on the boundary. Indeed, the BdG energy spectrum obtained for a long cylinder with open boundary condition along the $y$-direction, shown in Fig.~\ref{current}(a), reveals a pair of chiral edge states in the SC gap localized at each end.

It has been widely discussed whether such topological edge states can carry a physical charge current~\cite{2000Green,2005Wang,2015Kallin,2018Kallin,2024Franz}. 
We thus calculate the expectation value of the electron current operator ${\bm j}_{\alpha\beta}= -t\,\text{Im}\big[\langle c_{\bm{r}\alpha\sigma}^\dagger c_{\bm{r}^\prime\beta\sigma}\rangle\big]$, where $\alpha$ and $\beta$ denotes the sublattices of a pair of nn sites defining a bond. The current $\hat{\bm{j}}_{\alpha\beta}$ flows along the bond direction and is located at the bond center $(\bm{r}_\alpha+\bm{r}_\beta$)/2. We calculate the current distribution along the boundary direction under both open and periodic boundary conditions. As shown in Fig.~\ref{current}(b,c), in contrast to the case of periodic boundary conditions (blue) where persistent staggered supercurrent flows in the entire bulk, the open cylinder hosts counter-propagating charge current at the opposite edges carried by the topological chiral edge modes. This remarkable topological boundary charge current in the roton superconductor is due to the presence of spontaneously generated current vertex associated with the LC order in Eq.~\eqref{Hamil_cdw} in the CDW state.

It is instructive to discuss what determines the edge current flow direction. As explained in the previous sections, the chirality of the (anti-)roton SC ground state is described by the pairing angular momentum $L=\pm2$, which is controlled by the Josephson phase $\varphi$ according to the condition $L\varphi>0$. For a fixed LC configuration in the CDW state, the SC solutions of different chirality, i.e. different $L$ as shown in Fig.~\ref{instability} under different pairing interactions and complex CDW parameters, maintain the same topological invariant determined by the total Chern number in the Bogoliubov quasiparticle spectrum. Consequently, the direction of the topological edge current obtained above is determined by the TRS breaking normal state LC alone, regardless of the chirality of the SC state. The direction of the edge current flow would flip when the LC direction flips, say across the different $Z_2$ TRS breaking domains. These are unique features of the chiral topological SC state emerging from pairing in the partially filled Chern band of a Chern metal. 

In realistic materials, the edge current flows along the boundary of different $Z_2$ domains of the TRS breaking LC, which correspond to domains of the resulting roton SC states with different chiralities. 
This provides a plausible mechanism for the recently observed zero magnetic field SC diode effect and the critical current oscillations in applied magnetic fields in kagome superconductors CsV$_3$Sb$_5$~\cite{2024Le,2025Wang}. Specifically, the measured nonreciprocal critical currents in zero magnetic field requires spontaneous TRS breaking chiral domains with domain wall boundary currents~\cite{2025Varma}, in addition to the less difficult condition for inversion symmetry breaking. Remarkably, the polarity of the diode effect measured in zero-field in the SC state can be trained and flipped by flipping the direction of the training magnetic field in the normal state~\cite{2025Wang}. This is consistent with LC order in the normal state and roton superconductivity arising from pairing in the partially filled Chern band of the LC Chern metal.

\begin{figure*}
    \centering
    \includegraphics[width=\textwidth]{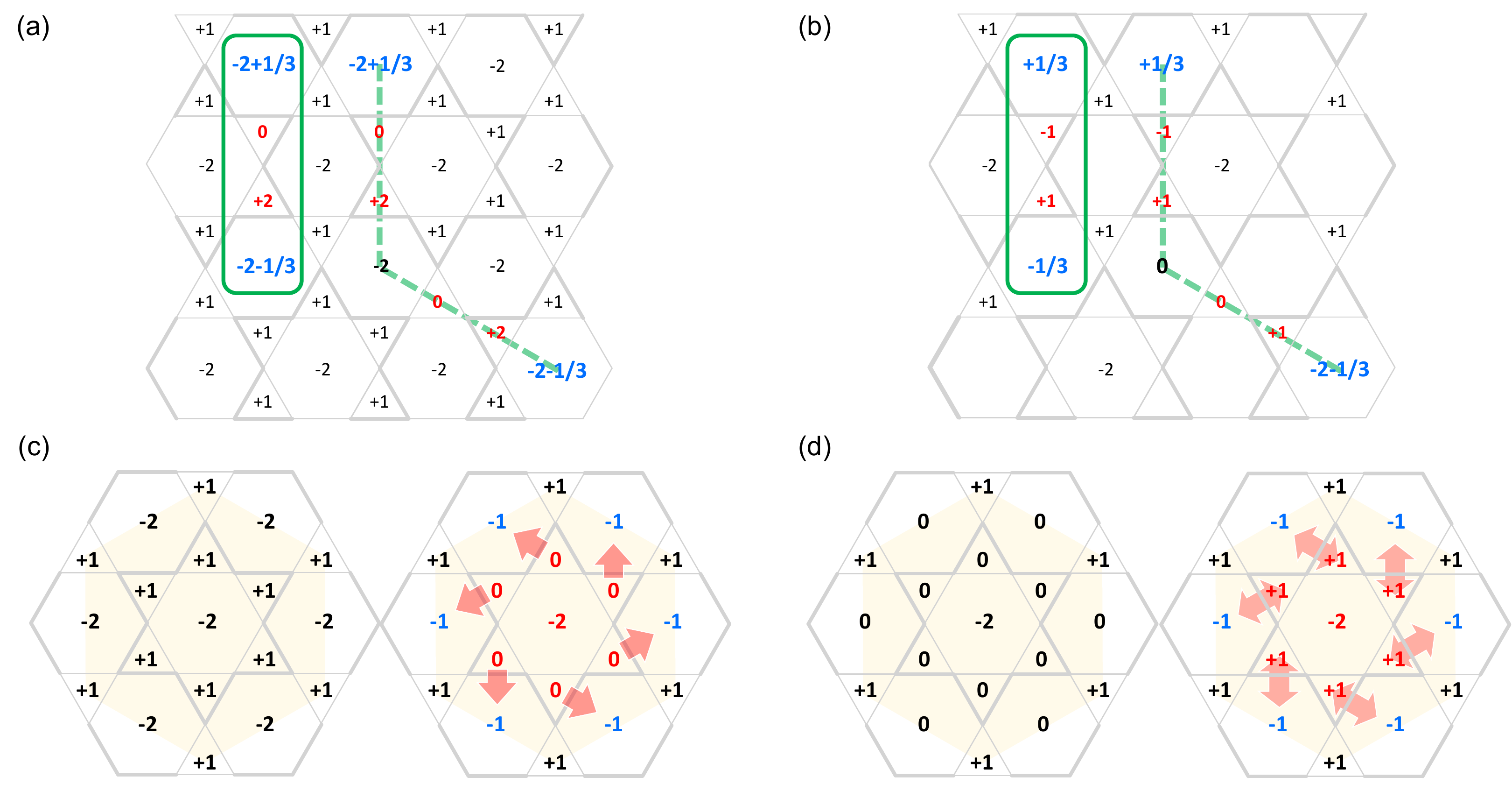}
    \caption{{\bf Fractional vortex pair excitations.} The creation of $\pm{1\over3}$ V-AV pair in the  $1\times1$ (a) and $2\times2$ (b) V-AV lattice of the roton SC state. The vorticity of the original V-AV lattice is marked numerically in black at the center of the hexagons and triangles. 
    A pair of fractional vortices, $(-2+1/3,\,-2-1/3)$ in (a) and $(1/3,-1/3)$ in (b) (labeled in blue) inside the green rectangle is created by a single phase slip of the two triangles in the middle, whose vorticity after the phase slip is labeled in red. The fractional V-AV pairs can be spatially separated by multiple phase slips along the green dashed line marked on the right in each case.
    (c) Creation of a $2\times2$ defect in a $1\times1$ roton state through $6$ phase slips between the hexagons and the contingent triangles indicated by red arrows. The $2\times2$ vortex defect is marked in red, which is separated from the $1\times1$ background by a vortex domain wall (marked in blue). (d) Creation of a $1\times1$ defect in a $2\times2$ roton state. The $1\times1$ defect is labeled in red, which is separated from the $2\times2$ background by a vortex domain wall (marked in blue).}
    \label{frac_vortex}
\end{figure*}

\section{Chiral phase fluctuations, fractional vortices, and charge-6${e}$ superconductivity}

We next turn to the fluctuation effects above the mean-field ground state of the roton superconductor.
It was argued recently that strong relative phase fluctuations of pairing on the Fermi pockets can produce an extended SC fluctuation region, where the SC state with charge-6e flux quantization can emerge~\cite{2023varma-wang}.
We are thus motivated to study the low-energy physics governed by the fluctuations of the relative chiral phases $\phi_j$ between the Cooper pairs on the CFPs in the roton superconductor described by the wave function in Eq.~\eqref{chiral-wf}. The fluctuations in the pairing amplitudes and the overall SC phase are assumed to be small and not affecting the qualitative physics discussed here.

The spatial distribution of the $120^\circ$ oriented internal phase, e.g. in Fig.~\ref{vortex_nn}(b), of the roton SC order parameter $\Delta(\bm{r}_\alpha)$ in Eq.~\eqref{op} suggests an analogy to the frustrated XY model on the kagome lattice~\cite{1992Huse,1992Harris,1993Coleman,2002Korshunov,2023Zhangguangming}. The vector spin chirality~\cite{1992Huse,2002Korshunov,2005Lee} in the language of the XY model corresponds to the vorticity in the relative Josephson phases of the superconductor. However, there are crucial differences. Since the TRS is already broken by the LC order in the CDW state at a higher energy scale than superconductivity, the chiral roton and antiroton states $\Psi_\pm$, i.e. the two chiral $120^\circ$ states with opposite vector spin chirality, are no longer degenerate, as can be seen in the energies $E_\pm$ above. Thus, the chirality fluctuations via the proliferation of the zero-energy domain wall separating degenerate chiral domains, crucial for the destruction of the $120^\circ$ order in the frustrated XY model on the kagome lattice~\cite{1992Huse,2001Huse,2002Korshunov,2023Zhangguangming}, are suppressed in the low-energy fluctuations of the roton superconductor. In order to select a roton state with a fixed chirality, appropriated anisotropy terms need to be included for explicit $Z_2$ symmetry breaking~\cite{sturla-2020prb,2023Castellani}. Moreover, the emergent $1\times1$ and $2\times2$ V-AV lattices (Fig.~\ref{vortex_nn}(b-c)) indicate that the roton superconductor is a textured SC state intertwined with pair density modulations at both the fundamental and the CDW reciprocal lattice wavevectors.

\subsection{Phase slips and fractional vortices}

The important low-energy fluctuations of the roton superconductor are associated with the change in the loop supercurrent configuration, akin to the supercurrent phase slips studied in Josephson junction kagome wire networks in an applied frustrating magnetic field~\cite{1997Rzchowski,2001Huse}. The quantum as well as thermally induced supercurrent phase slips cause the phase winding of the local SC order parameter to change abruptly by an integer multiple of $2\pi$, 
creating motion of the vortices and antivortices. To examine its effect, consider local phase slips at a pair of up and down triangles in the $1\times1$ roton state in Fig.~\ref{vortex_nn}(b). As shown in Fig.~\ref{frac_vortex}(a), where the vorticities of the V-AV lattice is labeled numerically, the supercurrent pattern changes as the vortex in the up triangle moves into the down triangle under current conservation, altering the vortex pattern from $(+1,+1)$ to $(0,+2)$. Two important consequences are in order. First, inside the green rectangle in Fig.~\ref{frac_vortex}(a), the local phase slip generates a pair of fractional ${1\over3}$ vortex and antivortex in the top and bottom hexagons, turning their vorticities from $(-2,-2)$ to $(-2+{1\over3},-2-{1\over3})$. Intriguingly, the fractional vortices can be separated by any distance following a sequence of such phase slips across the up and down triangles as indicted by the green dashed line in Fig.~\ref{frac_vortex}(a), creating isolated fractional $\pm{1\over3}$ V-AV pairs. For the $2\times2$ roton state, the fractional $\pm{1\over3}$ vortices can be generated in a similar fashion by phase slips, as illustrated in Fig.~\ref{frac_vortex}(b). 

The isolated fractional vortex and antivortex pair excitation located at the ends of line defects in the background V-AV lattice indicated by green dashed lines in Fig.~\ref{frac_vortex}(a-b) is one of the most essential properties of the roton superconductor. It offers a unique opportunity for its experimental detection. While the magnetic flux of the background V-AV lattice is difficult to measure in the ground state of a roton superconductor, the predicted spatially-separated fractional $\pm{1\over3}$ vortex and antivortex excitations 
should in principle be detectable by scanning SQUID at low temperatures.

Second, the $1\times1$ and the $2\times2$ V-AV lattices supporting pair density modulations are locally convertible by a sequence of multiple phase slips.
As shown in Fig.~\ref{frac_vortex}(c,d), the phase slips between the hexagons and the contingent triangles leads to the annihilation of vortices and antivortices. The local $1\times1$ V-AV configuration and supercurrrent pattern are turned into those surrounding the star-of-David in the $2\times2$ V-AV lattice and vice versa, reflecting the connection between the two spatial components by an ordered arrangement of phase slips.

\subsection{Charge-6e pairing and flux quantization}

The proliferation of the ${1\over3}$ fractional vortices through phase slips and the accompanied vortex motion
will lead to the melting of the hexatic V-AV lattice at higher temperatures above the zero resistance state. The latter can be described in the general framework of the KTHNY theory~\cite{1973KT,1979NH,1979Young}. The V-AV lattice melting in generalized chiral XY models mapped onto the two-dimensional Coulomb gas has been studied recently by renormalization group and Monte Carlo simulations~\cite{sturla-2020prb,2023Castellani}.
While more theoretical studies and numerical simulations are necessary to establish this picture for the roton superconductor, which is beyond the scope of the current work, the physics outlined here provides a simple picture for the possible emergence of a charge-6e SC state following the melting of the V-AV lattice that holds together the composite charge-2e roton superconductor.

Consider the chiral phase factors $\omega_j=e^{i\phi_j}$ in the wave function in Eq.~\eqref{chiral-wf} for the charge-2e roton superconductor with a fixed chirality. In the ground state, the relative phases between the Cooper pairs on the three CFPs are locked at $2\pi/3$, i.e. $\phi_{j}-\phi_{j+1}=2\pi/3$. Above the ground state, the internal phases $\phi_j$ fluctuate strongly. Since the sum of the internal phases is a part of and absorbed into the overall U(1) SC phase, the relative phase fluctuations have an important constraint $\sum_j\phi_j=0\mod{2\pi}$.
The internal phase fluctuations due to phase slips and vortex motion have a direct impact on the (quasi-)long-range order of the charge-2e roton superconductivity. This can be readily seen in the charge-2e SC order parameter in Eq.~\eqref{SC-2e} evaluated using the roton wave function in Eq.~\eqref{chiral-wf}. Clearly, the charge-2e order parameter is proportional to $e^{i\phi_j}$. Under strong relative chiral phase fluctuations, the correlation function of $e^{i\phi_j}$ decays exponentially, i.e. $\langle e^{i\phi_j}\rangle\to0$, such that the charge-2e SC order is destroyed. Similarly, the charge-4e order parameters also couple to and are thus suppressed by the strong relative phase fluctuations.

\begin{figure}
    \centering
    \includegraphics[width=0.7\columnwidth]{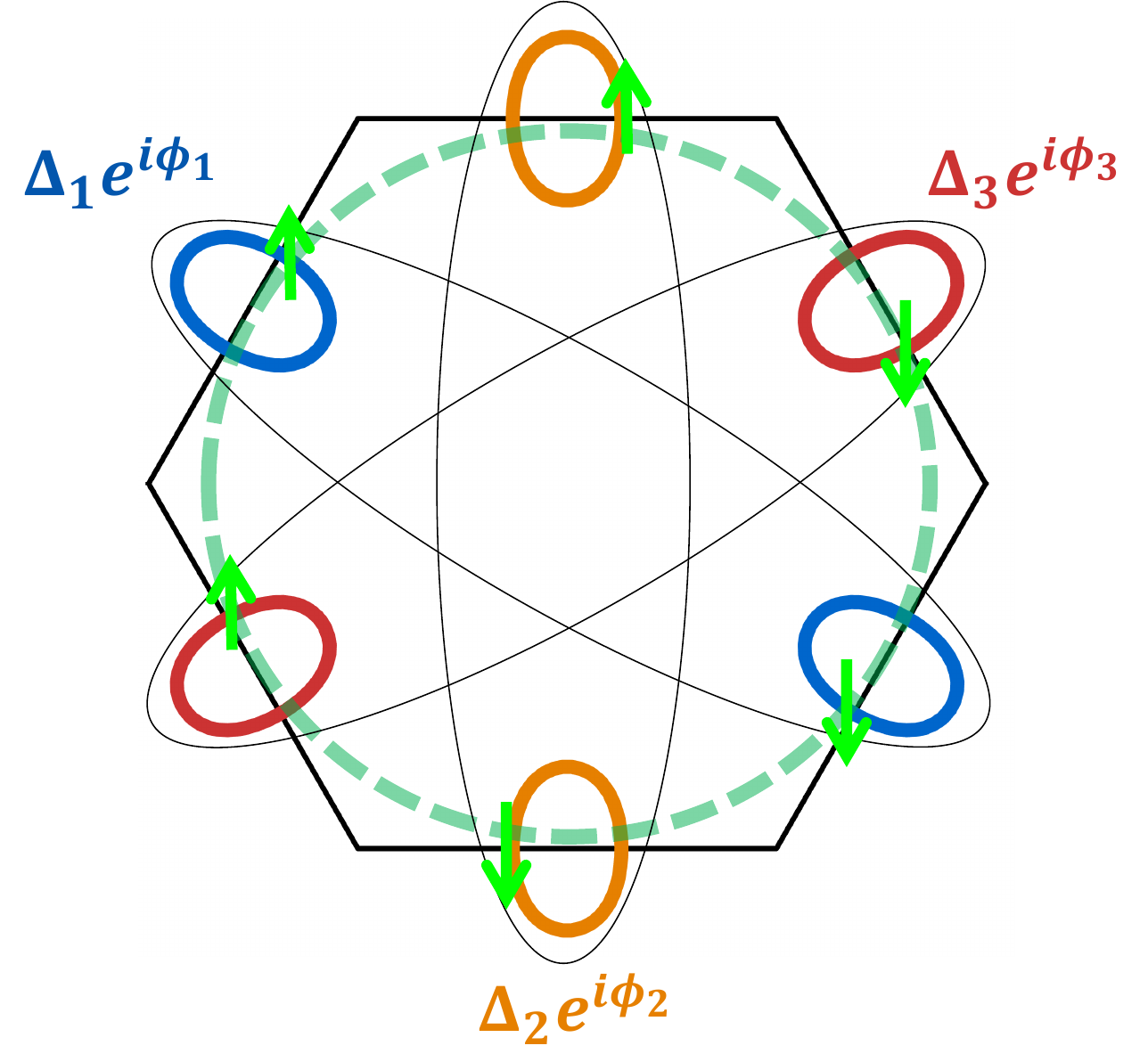}
    \caption{{\bf Schematic illustration of the 6e bound state.} The 6e bound state, as illustrated by the green dashed circle, consists of 3 Cooper pairs, each from pairing on one Fermi pocket and carries an relative phase $\phi_1,\phi_2,\phi_3$, respectively. The 6e bound state couples to the total phase $\phi_1+\phi_2+\phi_3=0 \mod{2\pi}$, and is thus immune to the internal phase fluctuations.}
    \label{6e-schematic}
\end{figure}

Remarkably, a different situation arises in the charge-6e order parameters, which can be evaluated from the roton wave function $\vert\Psi_{\rm roton}\rangle$ in Eq.~\eqref{chiral-wf},
\begin{widetext}
\begin{align}
    \Delta_{\rm 6e}({\bm r})&=\langle\Psi_{\rm roton}\vert c_{{\bm r}1} c_{{\bm r}1}
    c_{{\bm r}2} c_{{\bm r}2}
    c_{{\bm r}3} c_{{\bm r}3}\vert\Psi_{\rm roton}\rangle = {e^{i3\theta}\over 2^{3/2}}\sum_{\bk_1\bk_2\bk_3}   e^{i(\phi_{j_1}+\phi_{j_2}+\phi_{j_3})}
     \frac{g_{\bk_1}\psi_{11}^{{\bk_1}}}{1+|g_{\bk_1}|^2} \frac{g_{\bk_2}\psi_{22}^{{\bk_2}}}{1+|g_{\bk_2}|^2} \frac{g_{\bk_3}\psi_{33}^{{\bk_3}}}{1+|g_{\bk_3}|^2},
     \label{SC-6e}
\end{align}
\end{widetext}
where $j_i\in(1,2,3)$ labels the CFP that $\bk_i$ resides and the corresponding Cooper pair carries a phase $e^{i\phi_{j_i}}$. We have also restored the overall U(1) phase and $e^{i3\theta}$ signifies that the order parameter involves six electrons or three Cooper pairs. There are two distinct types of contributions to the charge-6e order parameter. One type is where the three Cooper pairs reside on either one or two CFPs and couple to a net chiral phase in Eq.~\eqref{SC-6e} and thus vanishes in the presence of strong chiral phase fluctuations, in a similar fashion as the 2e and 4e order parameter discussed above. However, there is another, intriguing type of 6e contributions where each of the three Cooper pairs resides on a different CFP, i.e. $j_1\ne j_2\ne j_3$ such that $e^{i(\phi_{j_1}+\phi_{j_2}+\phi_{j_3})}=1$, as illustrated schematically in Fig.~\ref{6e-schematic}. They are therefore completely decoupled from and immune to all internal chiral phase fluctuations. The latter allows the charge-6e order parameter $\Delta_{6e}$ and the corresponding higher harmonics $\Delta_{n6e}$ to survive strong chiral phase fluctuations that destroy the charge-2e and charge-4e order. 

This is precisely the microscopic mechanism behind the vestigial ordered higher-charge SC state proposed for the staged melting of a superconductor with a composite order parameter such as pair density wave superconductors~\cite{2009Kivelson,2011Tsunetsugu,2022Zhoub}, nematic superconductors~\cite{2021Fu,2021Yaoa}, multicomponent superconductors~\cite{2010Sudbo}, and in this work the hexatic roton superconductor with six-fold orientational order. We did not consider here the higher-charge superconductivity as vestigial order of the pair density wave order with an emergent V-AV lattice~\cite{2022Zhoub}, which was proposed following the STM observation of the $\frac{4}{3}a_0\times \frac{4}{3}a_0$ pair density wave in CsV$_3$Sb$_5$~\cite{2021Chen}. 
This is because the latter amounts to small SC modulation amplitudes on top of the large background of translation invariant SC condensate that is identified here as coming from the chiral roton state. We therefore focused on the melting of dominant hexatic roton superconductor due to internal chiral phase fluctuations.

We have thus obtained the smallest charged bound state, i.e. the charge-6e bound state formed by three Cooper pairs residing on the three CFPs as illustrated in Fig.~\ref{6e-schematic}, in the SC phase fluctuating regime. 
If the correlation length of the charge-6e state exceeds the perimeter of a ring structure, we expect the $hc/6e$ flux quantization to appear in the magnetoresistance oscillations, as observed in thin film CsV$_3$Sb$_5$ ring devices~\cite{2024Ge}. It is important to point out~\cite{2023varma-wang} that the charge-6e state in the experiments has not condensed since the $hc/6e$ flux quantization has only been observed in the wide resistive SC fluctuation regime below the $T_c$ onset. Thus the charge-6e superconductivity has only been observed as a mesoscopic phenomenon, which means that the charge-6e bound states are coherent on the scale of the perimeter of the ring devices as large as a micron~\cite{2024Ge}.

Theoretically, the charge-6e SC state, i.e. the quasi-long-range order of the 6e state with power-law decay correlations, can be reached in principle below its Kosterlitz-Thouless (KT) transition in two dimensions via the binding of thermally excited ${1\over3}$ fractional vortices
associated with the fluctuations in the global U(1) SC phase factor $e^{i3\theta}$ in Eq.~\eqref{SC-6e}. The latter are different and should not be confused with the $\pm{1\over3}$ fractional vortices of the internal Josephson phase associated with the charge-2e roton state. Based on the above discussions, we can indeed write down the coherent state wave function for the condensate of the charge-6e state
\begin{align}
 |\Psi_{\text{6e}}\rangle=&{\cal N}_{6e}\prod_{\bk_1\bk_2\bk_3}
    \bigl(1+e^{i3\theta}g_{\bk_1}g_{\bk_2} g_{\bk_3}b_{\bk_1}^{\dagger}
    b_{\bk_2}^{\dagger}b_{\bk_3}^{\dagger}\bigr)\vert \text{vac}\rangle \notag \\
    =&
    {\cal  N}_{6e}e^{\sum_{\bk_1\bk_2\bk_3} e^{i3\theta} g_{\bk_1}g_{\bk_2}g_{\bk_3} b_{\bk_1}^{\dagger}b_{\bk_2}^{\dagger}b_{\bk_3}^{\dagger}} |\text{vac}\rangle,
    \label{6e-coherent}
\end{align}
where $\bk_{1,2,3}$ belong to different CFPs in accord with the bound states illustrated in Fig.~\ref{6e-schematic}, and ${\cal N}_{6e}$ is the normalization factor given by ${\cal N}_{6e}^{-2}=\sum_n n!\big (\sum_{\bk_1\bk_2\bk_3}|g_{\bk_1}g_{\bk_2}g_{\bk_3}|^2\big )^n$.
It is important to note that $\vert\Psi_{6e}\rangle$ is not an eigenstate of the BCS-like pairing Hamiltonian and the higher-charge superconductivity does not have a mean field description. The microscopic conditions for realizing the higher-charge condensate are unclear currently. There are at least two possibilities for why a charge-6e zero-resistance state has not been observed. One is that the KT transition temperature of the 6e state, which is proportional to the interaction strength between $hc/6e$ vortices and antivortices, may be too low~\cite{2002Korshunov} and preempted by the charge-2e condensation at low-temperatures. The other is that the charge-6e state may decay into other incoherent states~\cite{2023varma-wang}. Future studies are necessary and desirable for understanding these extraordinary SC phenomena in the kagome superconductors. 

\section{Summary}

When dilute carriers are introduced into a Chern insulator, a Chern metal arises with small CFPs. The stable correlated phases deriving from the partially filled Chern band is central to the understanding of correlated and topological electronic matter.
In the context of a one-orbital effective model for the loop-current CDW in the kagome metals near vH filling~\cite{2022Zhoub}, we presented a theory for the emergence of an unprecedented topological roton superconductor through Cooper pairing on the three CFPs located at the CDW zone boundary of the Chern metal.

The hallmark of the roton superconductor is the condensation of multicomponent Cooper pairs carrying loop supercurrents and forming a hexagonal V-AV lattice. Thus the roton superconductor 
exhibits a vortex density wave order. The staggered nature of the persistent supercurrents escapes the Bloch theorem that forbids a uniform electric current in the ground state of a superconductor~\cite{1949Bohm}. Moreover, the staggered vortices with large momentum on the atomic scale can enter the superconductor without interfering with the Meissner screening, in analogy to the coexistence of superconductivity and antiferromagnetic order. Thus the roton superconductor exhibits the Meissner effect in an applied external magnetic field.

We showed that on the kagome lattice close to vH filling, the roton superconductor can emerge from a partially filled Chern band in the  $2\times2$ bond CDW state with loop-current order - a Chern metal normal state hosting three CFPs at the three inequivalent M valleys in the reconstructed zone. Quantum geometry plays a crucial role, because the phase of the complex Josephson coupling between the Cooper pairs on different CFPs is determined by the geometric Berry phase contribution associated with the discrete rotation or sublattice permutation. The large Josephson phase, arising even under small loop-current order, can drive the roton condensation forming the V-AV lattice.

Despite being an oversimplification when applied to the multiorbital kagome ``135'' materials,
the roton superconductor derived from the Chern metal normal state in the one-orbital kagome lattice model captures some of the most extraordinary features of the kagome superconductors. In addition to the TRS breaking supported by evidence from many experiments~\cite{2022Guguchia, 2024Le, 2024Yin, 2024Yinb}, the roton superconductor exhibits an anisotropic SC gap, which is consistent with the two sets of SC gaps observed by the STM tunneling spectra, including the anisotropic gap attributed to the V $d$-orbitals and the isotropic gap to the Sb $p$-orbital not included in our one-orbital model~\cite{zywang-nc,mode-nc2024,2024Nowack}. The TRS breaking roton superconductor is topological, exhibiting current-carrying chiral edge states. These properties can account for the zero-field spontaneous SC diode effect and the anomalous quantum oscillations attributed to current-carrying domain wall states observed in CsV$_3$Sb$_5$~\cite{2024Le}. The roton SC chiral domain walls are associated with those of the $Z_2$ TRS breaking LC order in the normal state, which provides theoretical support for the observed magnetic field training in the polarity of the nonreciprocal critical currents ~\cite{2025Wang}.

We showed that the V-AV lattice coexists and carries the momentum of the $2\times2$ CDW order in the LC Chern metal normal state. As a result, the order parameter of the roton superconductor exhibits intra-unit-cell chiral $2\times2$ and $1\times1$ pair density modulations, which have been observed by recent STM experiments~\cite{2024Yin,2025ligeng}. The translation symmetry breaking in a roton superconductor can also happen spontaneously, in which case the rotons and the associated V-AV lattice can condense at an independent momentum, leading to a primary pair density wave state~\cite{2022Zhoub}. We did not study the pair density wave formation here, which has been observed at $3\over4$ of the reciprocal lattice vector ${\bf G}$ by STM in  CsV$_3$Sb$_5$~\cite{2021Chen} and at ${1\over4}{\bf G}$ in the emergent SC state when the Cs atoms on the surface form an antiphase boundary for the stacking of the bulk CDW order~\cite{chenhui-surface}.
Such studies require microscopic consideration of the electronic interactions that favor translation symmetry breaking, finite-momentum pairing between different CFPs outside of scope of this work and are thus left for the future.

We argued that the hexatic charge-2e roton ground state becomes unstable due to supercurrent phase slips and fractional V-AV excitations in an extended fluctuating superconductivity regime created by the strong internal chiral phase fluctuations on the kagome lattice. A vestigial isotropic charge-6e state, a bound state of three Cooper pairs on the three different CFPs decoupled from internal phase fluctuations, can emerge with power-law SC correlations. This provides a plausible explanation for the experimental evidence for the charge-6e flux quantization in the extended fluctuation regime of thin-film CsV$_3$Sb$_5$ ring devices~\cite{2024Ge}.

The physics discussed here is also relevant for twisted bilayer and rhombohedral graphene systems where superconductivity has been observed in close proximity to orbital-driven quantum anomalous Hall states~\cite{2018Sua, 2020Bultinck, 2021Wang, 2021Liua,2024Jusc}. The emergence of the roton superconductor from a Chern metal normal state exemplifies a path toward intrinsic topological SC states by doping carriers into gapped topological phases of matter. In addition to capturing some of the essential properties of the kagome metals $A$V$_3$Sb$_5$, our findings based on the effective one-orbital model may have broader merit of its own.

\section{Acknowledgement}
    We thank Chandra Varma and Sen Zhou for invaluable discussions and Jun Ge, Jian Wang, Hui Chen, and Hongjun Gao for sharing their experimental insights. This work is completed while ZW is on sabbatical leave at the Kavli Institute for Theoretical Sciences (KITS), Chinese Academy of Sciences. ZW thanks the KITS for hospitality. This work is supported by the U.S. Department of Energy, Basic Energy Sciences Grant DE-FG02-99ER45747 and by Research Corporation for Science Advancement under Cottrell SEED Award No. 27856.

\bibliography{main.bib}

\begin{thebibliography}{99}
\bibitem [{\citenamefont {Li}\ and\ \citenamefont {et~al}()}]{2025ligeng}%
  \BibitemOpen
  \bibfield  {author} {\bibinfo {author} {\bibfnamefont {G.}~\bibnamefont {Li}}\ and\ \bibinfo {author} {\bibnamefont {et~al}},\ }\href@noop {} {\bibinfo  {journal} {To be published}\ }\BibitemShut {NoStop}%
\end{thebibliography}

\clearpage
\appendix
\section{Gauge Fixing}\label{gauge}
In the main text, we mentioned applying a unitary transformation $U_{\bm{k}}$ with matrix element $u_{\alpha n}^{\bm{k}}$ to obtain the energy dispersion $E_{n\bm{k}}$ and quasi-particle operator $f_{n\bm{k}}$:
\begin{equation}
	c_{\bm{k}\alpha\sigma}=\sum_{n\bk} u_{\alpha n}^{\bm{k}}f_{n\bm{k}\sigma},
\end{equation}
where columns of $U_{\bm{k}}$ corresponds to the energy eigen state of $H_{\text{K}} + H_{\text{CDW}}$, denoted as $|u_{n}^{\bm{k}}\rangle$. The effective interaction in the pairing channel is projected onto the CFPs using the pair wave function $|\psi_{\alpha\beta}^\bk\rangle$ defined in terms of $|u_{n}^{\bm{k}}\rangle$, as in Eq.~\eqref{int-proj}. Therefore, it is necessary to address the gauge degree of freedom in writing $|u_{n}^{\bm{k}}\rangle$. The gauge degree of freedom corresponds to the gauge transformation defined as:
\begin{equation}
    |u_n^{\bm{k}}\rangle\rightarrow |u_n^{\bm{k}}\rangle'=e^{i\tilde{\phi}_{\bm{k}}}
    |u_n^{\bm{k}}\rangle.
\end{equation}
Here both $|u_n^{\bm{k}}\rangle'$ and $|u_n^{\bm{k}}\rangle$ are physically identical energy eigenstates with the same eigenvalue $E_{n\bm{k}}$. We then fix the gauge according to the $C_6$ symmetry of the lattice. The gauge fixing is realized by
\begin{equation}
	|u^{\bm{k}}_n\rangle=e^{-i\tilde{\varphi}_{\alpha_j,\bm{k}}}|\tilde{u}^{\bm{k}}_n\rangle, \quad \bm{k}\in j.
\end{equation}
Here $|\tilde{u}^{\bm{k}}_n\rangle$ is the wave function under an arbitrary gauge (e.g., obtained through matrix diagonalization by a computer program), and $\tilde\varphi_{\alpha_j,\bm{k}}$ is the complex angle of $\tilde{u}_{\alpha_j,n}^{\bm{k}}$, the $\alpha_j$ component of $|\tilde{u}^{\bm{k}}_n\rangle$. In this way, we get $|u^{\bm{k}}_n\rangle$ as the wave function after gauge fixing. In the symmetric gauge, the sublattice index $\alpha_j$ is chosen following rotation symmetry. Specifically, for two pockets $(jj^\prime)$ related by one rotation operation, that is $M^{\text{R}}_{j^\prime}=\bm{R}_6(M^{\text{R}}_j)$, the corresponding sublattice indices used in the gauge fixing follow the rotation relation in real space, that is $\alpha_{j^\prime}=\bm{R}_6(\alpha_j)$. Here $\bm{R}_6$ denotes a six-fold rotation in $\bm{k}$-space around the $\Gamma$ point or in real space around the center of the SD.

\section{Partial Wave Expansion}\label{partial}
In the instability analysis, the SC order parameter $\Delta^f_j(\bk_j)$ is defined along each CFP. Here we write $\bk_j$ in terms of $\theta_\bk$, the polar angle around the $M^R_j$ point. The angular dependence of $\Delta^f_j(\theta_\bk)$ can be expressed through the partial wave expansion as
\begin{equation}
    \Delta^f_j(\theta_\bk)=\sum_l\Delta^f_{j,l}e^{il\theta_\bk}.
\end{equation}
Here $l=0,\pm2,\pm4,...$ denotes the angular momentum, which only takes even value due to spin-singlet (parity-even) pairing. The effective interaction $V_{jj^\prime}(\theta_\bk,\theta_{\bk^\prime})$ in the angular momentum channel reads:
\begin{equation}
    V^{ll^\prime}_{jj^\prime}= \sum_{\theta_\bk,\theta_{\bk^\prime}} V_{jj^\prime}(\theta_\bk,\theta_{\bk^\prime})e^{-il\theta_\bk}e^{il^\prime\theta_{\bk^\prime}}.
\end{equation}
In the main text, as an approximation, we consider the case of isotropic pairing along each CFP, which is equivalent to keeping only the $l=0$ component. The pairing is given by $\Delta^f_j(\theta_\bk)\approx\Delta^f_{j,0}\equiv\Delta^f_{j}$. The approximation is reasonable if the convergence of $V^{ll^\prime}$ over $l,l^\prime$ is fast. In practice, we found that the components from $l,l^\prime\ne0$ channel are at least an order of magnitude smaller than $V_{jj^\prime}^{00}$. Therefore, to the leading order, the pairing can be treated as isotropic along each pocket. The pairing order parameter can be written as a three-component vector $\bm{\Delta}=(\Delta^f_1,\Delta^f_2,\Delta^f_3)$. The free energy is given by Eq.~\eqref{freeE} in the main text.

\section{Discrete 3-fold Rotation and Sublattice Permutation}\label{permutation}
In this section, we show the discrete $\bk$-space rotation generated by $e^{-i\frac{2\pi}{3}\hat{L}_z}$ is equivalently generated by $\hat{P}_3$, the 3-fold sublattice permutation around the center of the SD.

The permutation operator denoted by $\hat{P}_3$ acts on the sublattice indices according to the clockwise 3-fold rotation around the center of SD. If we denote the initial sublattice index as $\alpha$ and the sublattice index after permutation as $[\alpha]$, the non-zero matrix element of the permutation operator under the sublattice basis is given by $P_{[\alpha]\alpha}=1$. For example, the element $P_{13}=1$ according to the sublattice indices in Fig.~\ref{basics}.

We first show that the full 3-fold rotation denoted by $\hat{R}_3=\hat{P}_3\otimes e^{i\frac{2\pi}{3}\hat{L}_z}$ is a symmetry of the system. Consider two $\bk$ points denoted by $\bk_{i,j}$  on two different CFPs, where $\bk_j$ can be generated from $\bk_i$ by a $2\pi/3$ clockwise rotation around the $\Gamma$ point. The corresponding rotation operator is denoted by $e^{i\frac{2\pi}{3}\hat{L}_z}$. The Hamiltonian, $H_{\bk_i}$ and $H_{\bk_j}$, as a function of $\bk$,  are therefore related by
\begin{equation}
    H_{\bk_j}=e^{iL_z2\pi/3} H_{\textbf{k}_i}e^{-iL_z2\pi/3}.
\end{equation}

To prove the full 3-fold rotation $\hat{R}_3$ is a symmetry of the system, i.e. $[\hat{R}_3,H_\bk]=0$ or $H_\bk=\hat{R}_3 H_\bk\hat{R}_3^{-1}$, consider one specific matrix element labeled $(H_{\bm{k}_i})_{\alpha\beta}$:
\begin{align}
    (H_{\bm{k}_i})_{\alpha\beta} &=t_{\alpha\beta}e^{i\bm{k}_i\cdot(\bm{r}_\alpha-\bm{r}_\beta)}\notag\\
    &=t_{[\alpha\beta]}e^{i\bm{k}_j\cdot(\bm{r}_{[\alpha]}-\bm{r}_{[\beta]})}\notag\\
    &=(H_{\bm{k}_j})_{[\alpha\beta]}\notag\\
    &=(\hat{P}_3e^{iL_z2\pi/3}  H_{\bm{k}_i}e^{-iL_z2\pi/3}\hat{P}_3^{-1})_{\alpha\beta}.
\end{align}
Here we use $[\alpha\beta]$ to denote the permutated bond connecting $[\alpha]$ and $[\beta]$. In the second equality, $t_{[\alpha\beta]}=t_{\alpha\beta}$ is from the rotation symmetry of the Hamiltonian. The exponential factor is the same because the bond rotation from $\bm{r}_\alpha-\bm{r}_\beta$ to $\bm{r}_{[\alpha]}-\bm{r}_{[\beta]}$ and the $\bm{k}$-rotation from $\bm{k}_i$ to $\bm{k}_j$ are performed in the same direction, leaving the vector product $\bk\cdot(\bm{r}_\alpha-\bm{r}_\beta)$ unchanged. The last equality follows from the definition of the $\bm{k}$-rotation and the permutation operator. Therefore, the operator $\hat{R}_3=\hat{P}\otimes e^{iL_z2\pi/3}$ commutes with $H_{\bm{k}}$ and is a symmetry operation.

The energy eigenstate can be categorized according to the eigen value of $\hat{R}_3$. Because $(\hat{R}_3)^3=1$, the eigen value of the operator can only take three values, namely $\{1, e^{i2\pi/3}, e^{-i2\pi/3}\}$. In our case, the gauge choice in the gauge-fixing section explicitly yields the eigenvalue to be $1$ for all the energy eigen states. This leads to:
\begin{equation}
    \hat{R}_3|u^{\textbf{k}_i}\rangle=|u^{\textbf{k}_i}\rangle.
\end{equation}
And for the pair wave functions, similar relation can be obtained:
\begin{equation}
    \hat{R}_3|\psi^{\textbf{k}_i}_m\rangle=|\psi^{\textbf{k}_i}_m\rangle.
\end{equation}
Strictly speaking, the rotation operator for the pair wave function should be the product of two single particle operators. We drop the product in the notation for convenience. In the end, we get:
\begin{align}
    |\psi^{\bk_j}_m\rangle=e^{-i\frac{2\pi}{3}\hat{L}_z}|\psi^{\bk_i}_m\rangle&=e^{-i\frac{2\pi}{3}\hat{L}_z}\hat{R}_3|\psi^{\bk_i}_m\rangle\notag\\
    &=\hat{P}_3|\psi^{\bk_i}_m\rangle.
    \label{permutation-k}
\end{align}
Therefore, the discrete rotation $e^{-i\frac{2\pi}{3}\hat{L}_z}$ and the permutation $\hat{P}_3$ are equivalent. In this way, Eq.~\eqref{rotation_phase} can be expressed as:
\begin{align}
     \langle\psi^{\bk_i}_m|\psi^{\bk_j}_m\rangle&=\langle\psi^{\bk_i}_m| e^{-i\frac{2\pi}{3}\hat{L}_z}|\psi^{\bk_i^\prime}_m\rangle = \langle\psi^{\bk_i}_m| \hat{P}_3|\psi^{\bk_i^\prime}_m\rangle.
\end{align}

\section{Self-Consistent Solutions}\label{self-consistent}
The self-consistent calculations are performed directly in real space. The singlet pairing meanfields are introduced as:
\begin{equation}
    \Delta_{\langle \alpha\beta\rangle_m}=\frac12\left[\langle c_{\bm{r}\alpha\downarrow} c_{\bm{r}'\beta\uparrow}\rangle -\langle c_{\bm{r}\alpha\uparrow} c_{\bm{r}'\beta\downarrow}\rangle\right].
\end{equation}
The kinetic part of the BdG Hamiltonian follows from Eq.~\eqref{tb-hopping}. The pairing part of the BdG Hamiltonian is given by:
\begin{align}
    H^{\text{MF}}_{\text{int}}=-W_m&\sum_{\langle \bm{r}\alpha;\bm{r}'\beta\rangle_m}[\Delta^*_{\langle \alpha\beta\rangle_m}(c_{\bm{r}\alpha\downarrow}c_{\bm{r}'\beta\uparrow}-c_{\bm{r}\alpha\uparrow}c_{\bm{r}'\beta\downarrow})\\
    &+\Delta_{\langle \alpha\beta\rangle_m}(c_{\bm{r}\alpha\uparrow}^\dagger c_{\bm{r}'\beta\downarrow}^\dagger-c_{\bm{r}\alpha\downarrow}^\dagger c_{\bm{r}'\beta\uparrow}^\dagger)]+\text{Const}.\notag
\end{align}
The BdG Hamiltonian can be written as:
\begin{equation}
    H_{\text{BdG}} = H_{\text{K}} + H_{\text{CDW}} + H_{\text{int}}^{\text{MF}}.
    \label{BdG-Hamil}
\end{equation}
Here $H_{\text{K}} + H_{\text{CDW}}$ is defined in the main texts. The BdG Hamiltonian can be diagonalized to give the BdG quasiparticle $\gamma_{n\bm{k}}$ and the associated dispersion:
\begin{equation}
    V_{\bm{k}}^\dagger H_{\text{BdG}} V_{\bm{k}} = \Lambda_{\bm{k}}^{\text{BdG}}.
\end{equation}
Here $\Lambda_{\bm{k}}^{\text{BdG}}$ is the diagonal matrix formed by the quasi-particle energy eigenvalues $E_{n\bm{k}}^{\text{BdG}}$. The transformation matrix $V_{\bm{k}}$ with matrix elements $v_{\alpha n}^{\bm{k}}$ can be written explicitly as:
\begin{align}
    c_{\bm{k}\alpha \uparrow}&=\sum_nv_{\alpha n}^{\bm{k}} \gamma_{n\bm{k}},\\
    c_{-\bm{k}\alpha \downarrow}^\dagger&=\sum_nv_{\alpha +12,n}^{\bm{k}} \gamma_{n\bm{k}},
\end{align}
where $12$ is the number of sublattices in the $2\times2$ unit cell. The self-consistent equations for $\Delta_{\langle \alpha\beta\rangle_m}$ is constructed using $V_{\bm{k}}$:
\begin{align}
    \Delta_{\langle \alpha\beta\rangle_m}=& \frac12\sum_n\Big[v_{\alpha+12, n}^{\bm{k}*} v_{\beta n}^{-\bm{k}*}e^{-i\bm{k}\cdot(\bm{r}_\alpha-\bm{r}_\beta)}n_F(E^{\text{BdG}}_{n\bm{k}})\\
    &-v_{\alpha n}^{\bm{k}} v_{\beta+12,n}^{-\bm{k}*}e^{i\bm{k}\cdot(\bm{r}_\alpha-\bm{r}_\beta)}\big(1-n_F(E^{\text{BdG}}_{n\bm{k}})\big)\Big].\notag
\end{align}
Here $n_F(E^{\text{BdG}}_{n\bm{k}})$ is the Fermi-Dirac distribution function for $E_{n\bm{k}}^{\text{BdG}}$. The SC order parameter can then be solved self-consistently.

The quasiparticle pairing order parameter $\Delta^f_j(\bk_j)$ can be obtained using $\Delta_{\langle\alpha\beta\rangle_m}$ based on the quasiparticle wave function in eq.~\eqref{uk}:
\begin{align}
    \Delta^f_j(\bk_j)&=\langle f_{-\bk_j\downarrow}f_{\bk_j\uparrow}\rangle\notag\\
    &=\sum_{\langle\alpha\beta\rangle_m}u_{\alpha n}^{-\bk_j*}u_{\beta n}^{\bk_j*}e^{i\bk_j\cdot(\bm{r}_\alpha-\bm{r}_\beta)}\Delta_{\langle\alpha\beta\rangle_m}.
\end{align}
Here $n$ refers to the band crossed by the Fermi level in the LC CDW dispersion.

\section{From internal chiral phases of Cooper pairs on CFPs to V-AV lattice in real space}\label{connection}

The 3-component roton superconductor with $L=\pm2$ can be understood as three phase-uniform pairing states on each of the three sublattices. In this section, we illustrate the connection between the internal chiral phase factor $\phi_j$ of the Cooper pairs on the CFPs and the chiral phase of the onsite pairing order parameter $\Delta(\bm{r}_\alpha)$ in real space. To demonstrate the relation, the permutation relation in Appendix.~\ref{permutation} is used. 

We consider first the pairing $\Delta^f_j(\bk_j)$ along the CFPs. In the instability analysis, the pairing order parameters $\Delta^f_j(\theta_{\bk_j})$ can either be in phase in the $(111)$ order or maintain an internal chiral phase among components in the $120^\circ$ order. The phase relation can be expressed as:
\begin{equation}
    \Delta_1^f(\theta_{\bk_1}): \Delta_2^f(\theta_{\bk_2}): \Delta_3^f(\theta_{\bk_3})=e^{i\phi_1}:e^{i\phi_2}:e^{i\phi_3}.
    \label{pairing_phase}
\end{equation}
Here $\theta_{\bk_j}$ denote the internal polar angle within the $j$th pocket. The phase associated with each CFP satisfies $\phi_{j}-\phi_{j+1}=\pm2\pi/3$ in the $120^\circ$ roton (anti-roton) state while $\phi_{j}-\phi_{j+1}=0$ in the $(111)$ state. Before moving forward, we write the $\bk_j$ points belonging to different CFP using the counter-clockwise 3-fold rotation around the $\Gamma$ point denoted by $e^{-i\frac{2\pi}{3}L_z}$. For example, $\bk_2$ here is obtained by one counter-clockwise 3-fold rotation operation of $\bk_1$ with $\theta=2\pi/3$, therefore we write $\bk_2=[\bk_1]$ with $[\bk]$ for the new coordinate after a 3-fold rotation.

Next, consider the onsite pairing order parameter $\Delta(\bm{r}_\alpha)$ defined in eq.~\eqref{po}. Because the order parameter $\Delta(\bk)$ is peaked near the Fermi surface, we write $\Delta(\bm{r}_\alpha)$ in terms of $\Delta_j^f(\bk_j)$ along the CFPs and the pair wave function $|\psi_m^{\bk_j}\rangle$ (for onsite pairing order parameter $\Delta(\bm{r}_\alpha)$, $m=0$ and the corresponding wave function components are denoted by $\psi_{\alpha\alpha}^{\bk_j}$):
\begin{align}
    \Delta(\bm{r}_\alpha)=\langle c_{\bm{r}\alpha\uparrow}c_{\bm{r}\alpha\downarrow}\rangle&=\sum_{j\bk_j}u_{\alpha n}^{\bk_j}u_{\alpha n}^{-\bk_j}\Delta^f_j(\theta_{\bk_j})\notag\\
    &=\frac1{\sqrt2}\sum_{j\bk_j}\psi_{\alpha\alpha}^{\bk_j}\Delta^f_{j}(\theta_{\bk_j}).
    \label{pairing-r2f}
\end{align}
The last equality follows from the definition of $\psi_{\alpha\alpha}^\bk$ below eq.~\eqref{int-proj}.

To illustrate the origin of the chiral phase associated with $\Delta(\bm{r}_\alpha)$, we consider two sublattices that are related by an $R_3$ rotation around the center of the SD. The sublattice index after rotation is denoted by $[\alpha]$ as in Appendix.~\ref{permutation}. For example, the sublattice index $2=[1]$ according to the labeling in Fig.~\ref{basics} (d). We have demonstrated in Appendix.~\ref{permutation} that the pair wave function satisfies $\psi_{\alpha\alpha}^{\bk_j}=\psi_{[\alpha\alpha]}^{[\bk_j]}$. The relation between the onsite pairing order parameter $\Delta(\bm{r}_\alpha)$ and $\Delta(\bm{r}_{[\alpha]})$ can be derived as follows:
\begin{align}
    \Delta(\bm{r}_{[\alpha]})
    &=\sum_{j\bk_j}\psi_{[\alpha\alpha]}^{\bk_j}\Delta^f_{j}(\theta_{\bk_j}) \notag\\ 
    &= \sum_{j\bk_j}\psi_{\alpha\alpha}^{[\bk_j]}\Delta^f_{j}(\theta_{\bk_j}) \notag\\ 
    &= \sum_{j\bk_j}\psi_{\alpha\alpha}^{\bk_j}\Delta^f_{[j]}(\theta_{[\bk_j]}) \notag\\ 
    &= \sum_{j\bk_j}e^{i(\phi_{[j]}-\phi_j)} 
    \psi_{\alpha\alpha}^{\bk_j}\Delta^f_{j}(\theta_{\bk_j}) \notag\\ 
    &\xlongequal{\text{roton}} e^{-i2\pi/3} \Delta(\bm{r}_\alpha)
\end{align}
Here the first equality is derived in eq.~\eqref{pairing-r2f}. The second equality follows eq.~\eqref{permutation-k}. We inter-change the sum over $\bk_j$ to $[\bk_j]$ and get the equality in the third line. The fourth equality follows from the phase relation of the pairing ground state, given by eq.~\eqref{pairing_phase}. Furthermore, because $\phi_{[j]}-\phi_j=\phi_{j+1}-\phi_j=\mp2\pi/3$ in the roton (anti-roton) solution is independent of the pocket index $j$, it can be taken out of the sum. As a result, in the $120^\circ$ state with $L=+2$ (roton), the phase difference between $\Delta(\bm{r}_\alpha)$ and $\Delta(\bm{r}_{[\alpha]})$ given by the last equality is in agreement with the vortex-antivortex lattice configuration shown in Fig.~\ref{vortex_nn} (b,c). We therefore demonstrated that in the roton state the relative phase between the sublattices related by $R_3$ can be written as $\Delta(\bm{r}_{[\alpha]}):\Delta(\bm{r}_{\alpha})=e^{-i2\pi/3}$, which originates from the relative internal phases $\phi_{j+1}-\phi_j$ among Cooper pairs on CFPs. In general, consider as an example the three sublattices indexed by $1,2,3$ according to Fig.~\ref{basics}(d), the ratio among the pairing order parameter can be written as 
\begin{equation}
    \Delta(\bm{r}_1):\Delta(\bm{r}_2):\Delta(\bm{r}_3)=e^{i\phi_1}:e^{i\phi_2}:e^{i\phi_3}.
    \label{ratio1}
\end{equation}

We next turn to the expression in eq.~\eqref{op}:
\begin{align}
    \Delta (\bm{r}_\alpha)=\sum_{\gamma=1}^3 e^{i\varphi_\gamma} \bigl[ &A_1 \cos \bm{G}_\gamma\cdot(\bm{r}_\alpha-\bm{r}_0)
    \notag\\
    &+A_2 \cos\bm{Q}_c^\gamma\cdot(\bm{r}_\alpha-\bm{r}_0+\delta_\gamma)\bigr].
\end{align}
and show that the phase factor associated with each component is directly reflected in the local pairing order parameter $\Delta(\bm{r}_\alpha)$ through superposition. To illustrate this, we begin by considering the $1\times1$ component with three wave vectors are denoted as (the $2\times2$ component corresponds to $\bm{Q}_c^\gamma=\bm{G}_\gamma/2$):
\begin{equation}
    \bm{G}_1=\frac{4\pi}{\sqrt3}(0, 1),\, \bm{G}_2=\frac{4\pi}{\sqrt3}(-\frac{\sqrt3}{2}, -\frac12),\, \bm{G}_3=\frac{4\pi}{\sqrt3}(\frac{\sqrt3}{2}, -\frac12).
\end{equation}
The center of the pair density modulation is chosen at $\bm{r}_0=(\frac14,\frac{3\sqrt3}{4})$, corresponding to the center of SD. We consider three rotation-related sublattices with their corresponding coordinates:
\begin{equation}
    \bm{r}_1=(\frac14,\frac{\sqrt{3}}4),\, \bm{r}_2=(0,0),\, \bm{r}_3=(\frac12,0).
\end{equation}
The modulations are therefore determined by the phase $\bm{G}_\gamma\cdot(\bm{r}_\alpha-\bm{r}_0)$, which are summarized in Table.~\ref{tab:phase}. The modulation phase of $\bm{Q}_c^\gamma\cdot(\bm{r}_\alpha-\bm{r}_0)$ is simply half the value of $\bm{G}_\gamma\cdot(\bm{r}_\alpha-\bm{r}_0)$.
\begin{table}[htbp]
    \centering
    \begin{tabular}{c|ccc}
         & $\bm{r}_1- \bm{r}_0$ & $\bm{r}_2- \bm{r}_0$ & $\bm{r}_3- \bm{r}_0$ \\
         \hline
        $\bm{G}_1$ & $-2\pi$ & $\pi$ & $\pi$ \\
        $\bm{G}_2$ & $\pi$ & $2\pi$ & $\pi$ \\
        $\bm{G}_2$ & $\pi$ & $\pi$ & $2\pi$ \\
        \hline
    \end{tabular}
    \caption{Phase of $\bm{G}_\gamma\cdot\bm{r}_\alpha$ in the density modulation.}
    \label{tab:phase}
\end{table}

The real space order parameter can then be expressed as:
\begin{align}
    \Delta(\bm{r}_1)&=A_1(e^{i\varphi_1} - e^{i\varphi_2} - e^{i\varphi_3})+A_2e^{i\varphi_1}\notag\\
    &=e^{i\varphi_1} \left(A_1(1 - e^{i(\varphi_2-\varphi_1)} - e^{i(\varphi_3-\varphi_1)})+A_2\right),\\
    \Delta(\bm{r}_2)&=A_1(e^{i\varphi_2} - e^{i\varphi_1} - e^{i\varphi_3})+A_2e^{i\varphi_2}\notag\\
    &=e^{i\varphi_2} \left(A_1(1 - e^{i(\varphi_1-\varphi_2)} - e^{i(\varphi_3-\varphi_2)})+A_2\right),\\
    \Delta(\bm{r}_3)&=A_1(e^{i\varphi_3} - e^{i\varphi_1} - e^{i\varphi_2})+A_2e^{i\varphi_2}\notag\\
    &=e^{i\varphi_3} \left(A_1(1 - e^{i(\varphi_1-\varphi_3)} - e^{i(\varphi_2-\varphi_3)})+A_2\right).
\end{align}
Note in all the (anti-)roton and the isotropic SC order, the phase $\varphi_\gamma$ maintain $\varphi_1-\varphi_2=\varphi_2-\varphi_3=\varphi_3-\varphi_1$ up to a $2\pi$ shift. In this way, the local pairing order parameter $\Delta(\bm{r}_\alpha)$ in the above expressions can be written in terms of a specific $e^{i\varphi_\gamma}$ related to the sublattice index multiplied by a common factor in the bracket. We can then obtain the ratio among $\Delta(\bm{r}_\alpha)$ as $\Delta(\bm{r}_1):\Delta(\bm{r}_2):\Delta(\bm{r}_3) = e^{i\varphi_1}:e^{i\varphi_2}:e^{i\varphi_3}$. Note it has been derived in eq.~\eqref{ratio1} that the same ratio is given by the internal phase $\phi_j$ of Cooper pairs on the CFPs. We therefore establish that the phase $\varphi_\gamma$ appearing in eq.~\eqref{op} and the internal phase $\phi_j$ are of the same origin.

\end{document}